# Probing fMRI brain connectivity and activity changes during emotion regulation by EEG neurofeedback

Amin Dehghani[a], Hamid Soltanian-Zadeh[a,b,c], Gholam-Ali Hossein-Zadeh[a,b,*]

[a] School of Electrical and Computer Engineering, University of Tehran, Tehran, Iran

[b] School of Cognitive Sciences, Institute for Research in Fundamental Sciences (IPM), Tehran, Iran

[c] Medical Image Analysis Lab., Departments of Radiology and Research Administration, Henry Ford Health System, Detroit, MI, USA

* Corresponding author: ghzadeh@ut.ac.ir

Address:

Gholam-Ali Hossein-Zadeh,

Scientific and Technical Deputy of Iranian National Brain Mapping Lab (NBML)

Prof. of School of Electrical and Computer Engineering

College of Engineering, University of Tehran

North Kargar Ave.

Tehran 14395-515, Iran

Tel: (+98)-21-8208-4178, (+98)-21-8822-5396

## Abstract

Neurofeedback is a non-invasive brain training with long-term medical and non-medical applications. Despite the existence of several emotion regulation studies using neurofeedback, further investigation is needed to understand interactions of the brain regions involved in the process. We implemented EEG neurofeedback with simultaneous fMRI using a modified happiness-inducing task through autobiographical memories to upregulate positive emotion.

The results showed increased activity of prefrontal, occipital, parietal, and limbic regions and increased functional connectivity between prefrontal, parietal, limbic system, and insula in the experimental group. New connectivity links were identified by comparing the functional connectivity of different experimental conditions within the experimental group and between the experimental and control groups.

The proposed multimodal approach quantified the changes in the brain activity (up to 1.9% increase) and connectivity (FDR-corrected for multiple comparison, q = 0.05) during emotion regulation in/between prefrontal, parietal, limbic, and insula regions. Psychometric assessments confirmed significant changes in positive and negative mood states by neurofeedback with a p-value smaller than 0.002 in the experimental group.

This study quantifies the effects of EEG neurofeedback in changing functional connectivity of all brain regions involved in emotion regulation. For the brain regions involved in emotion regulation, we found significant BOLD and functional connectivity increases due to neurofeedback in the experimental group but no learning effect was observed in the control group. The results reveal the neurobiological substrate of emotion regulation by the EEG neurofeedback and separate the effect of the neurofeedback and the recall of the autobiographical memories.

*Keywords*: Neurofeedback, Simultaneous recording of fMRI and EEG, emotion regulation, frontal asymmetry, functional connectivity, autobiographical memory.

## 1. Introduction

Neurofeedback is a non-invasive brain training procedure with various clinical applications. In neurofeedback, a feedback is provided to a subject based on his/her brain activity in order to self-regulate his/her brain function. EEG and fMRI are two main neuroimaging modalities used in the study of neurofeedback.

Emotion regulation consists of control and management of emotional states, mood, and affect with specific strategies, e.g., situation selection, situation modification, attentional deployment (like distraction and rumination), cognitive change (like reappraisal and distancing), and response modulation (like exercise) (Gross, 1998; Koole, 2009). The common strategies for emotion regulation can be categorized into implicit and explicit strategies based on conscious awareness.

Simultaneous recording of EEG and fMRI provides complementary information and allows for a more comprehensive understanding and research in neurofeedback by exploring EEG and fMRI correlation (EEG-informed fMRI), fusion analysis, and validation of effectiveness of the applied paradigm for a specific purpose. EEG-based neurofeedback has applications in the treatment of attention deficit/hyperactivity disorder (ADHD) (Lubar, Swartwood, Swartwood, & O'Donnell, 1995; Zuberer, Minder, Brandeis, & Drechsler, 2018), schizophrenia (Bolea et al., 2010), insomnia (Hammer, Colbert, Brown, & Ilioi, 2011), drug addiction (Lackner et al., 2016), autism (Coben, Linden, & Myers, 2010), epilepsy (Kaur & Singh, 2017; Linhartová et al., 2019; Saxby & Peniston, 1995; Walker & Kozlowski, 2005), anxiety (Mennella, Patron, & Palomba, 2017), pain (Kubik & Biedroń, 2013), eating disorders (Bartholdy, Musiat, Campbell, & Schmidt, 2013), Parkinson disease (Rossi-Izquierdo et al., 2013), obsessive compulsive disorder (Hammond, 2003), post-traumatic stress disorder (PTSD) (Gapen et al., 2016) in addition to other psychological applications like emotion regulation (Dennis & Solomon, 2010; Linhartová et al., 2019; Quaedflieg et al., 2015).

Most of the EEG neurofeedback protocols try to modulate the EEG signal amplitude or power in particular frequency bands. Recent development in MRI pulse sequences has made it possible to use fMRI for neurofeedback which is named real time fMRI (Cohen Kadosh et al., 2016; DeBettencourt, Cohen, Lee, Norman, & Turk-Browne, 2015; D. Y. Kim, Yoo, Tegethoff, Meinlschmidt, & Lee, 2015; Koush et al., 2013; Megumi, Yamashita, Kawato, & Imamizu, 2015; Sarkheil et al., 2015; Sherwood, Kane, Weisend, & Parker, 2016; Subramanian et al., 2011; Sulzer et al., 2013). EEG neurofeedback with simultaneous fMRI has been proposed in several studies, Theta/Alpha power ratio was used by (Kinreich, Podlipsky, Intrator, & Hendler, 2012) to enhance relaxation in subjects, which is useful in a range of clinical application such as PTSD and ADHD.

EEG frontal asymmetry has been used in several studies for emotion regulation based on the approach – withdrawal model proposed in (Davidson, Ekman, Saron, Senulis, & Friesen, 1990).

According to this model, withdrawal emotion or negative affect (such as fear, sadness, and disgust) associates with higher activity in the right hemisphere. Conversely, increasing activity in the left hemisphere associates with emotional states such as joy or anger (Davidson et al., 1990; Quaedflieg et al., 2015). Several applications of EEG frontal asymmetry are listed in (Coan & Allen, 2004). As mentioned in (J. J. B. Allen & Reznik, 2015; Coan & Allen, 2004; Mennella et al., 2017; Meyer et al., 2018; Peeters, Oehlen, Ronner, Van Os, & Lousberg, 2014), EEG frontal asymmetry can be used as an indication of different emotional states and a biomarker for PTSD, anxiety, and depression. In (Cavazza et al., 2014), EEG neurofeedback based on frontal alpha asymmetry with simultaneous fMRI was used based on interactive narrative paradigms. The success rate of emotion regulation and changing the frontal asymmetry in this study was low. In addition, this study did not include a control group.

According to a recently published review paper in emotion regulation using fMRI neurofeedback (Linhartová et al., 2019), several brain regions including amygdala, insula, anterior cingulate cortex, and prefrontal/frontal regions (dorsomedial prefrontal cortex, dorsolateral prefrontal cortex, and orbitofrontal cortex) are involved in emotion regulation and treatment of mental disorders such as depression, PTSD, and anxiety. In some previous studies (Koush et al., 2017; Vadim Zotev et al., 2011a), interactions among a small number of regions like amygdala and other regions were evaluated. Notwithstanding with several existing emotion regulation studies using neurofeedback or retrieval of autobiographical memories, understanding the interaction of the brain regions involved in emotion regulation through neurofeedback still needs further investigation by considering brain regions with key roles in emotion regulation.

In this study, we implement simultaneous fMRI–EEG recording during an EEG neurofeedback using an induced happiness task for emotion regulation through positive autobiographical memories. The fMRI signal changes, the EEG frontal asymmetry, the psychometric tests, and the brain activations are evaluated to detect changes in the activation of prefrontal and subcortical brain regions including limbic system and insula during neurofeedback by the proposed paradigm. The involvement of the prefrontal, limbic, and insular regions has been reported in several studies dealing with emotion regulation or retrieval autobiographical memories. Modulation of the activity and connectivity of the brain regions provides valuable information for studying the neural network related to emotion regulation. It can be also used to monitor the changes in behavior and cognition for the treatment of mental disorders such as PTSD and major depression disorder (MDD).

The aim of this study is to investigate how EEG neurofeedback changes the function and connectivity of the brain regions during emotion regulation and separate the effect of paradigm and neurofeedback during emotion regulation. We test the hypothesis that healthy participants can learn to control and voluntarily regulate the activity and connectivity in/between the prefrontal, limbic, and insular regions by means of EEG neurofeedback. The connectivity results can be used for the design of new neurofeedback paradigms for emotion regulation, especially using real time fMRI, as a therapeutic tool for the treatment of the mental disorders.

## 2. Methods

### 2.1 Task Design

The research protocol is approved by the ethics committees of the Iran University of Medical Sciences, Tehran, Iran. 18 healthy subjects (age 26.7 ± 3.6 years, all male) as the experimental group, and 14 healthy subjects (age 27 ± 3.8 years, all male) as the control group participated in this study. Participants in the control group are provided (without their knowledge) with sham EEG neurofeedback. The exclusion criteria are prior history of major psychiatric or neurological disorder, drug or alcohol abuse during the past year, brain surgery, and contraindications to MRI. Before the experiment, two psychometric tests including Beck's Depression Inventory (Craven et al., 2005) and General Health Questionnaire - 28 (GHQ-28) (Nazifi et al., 2014) are completed by each participant. The mean ± standard deviations of Beck’s Depression Inventory and GHQ-28 for all participants are 6.8 ± 3 and 2.2 ± 2, respectively. Therefore, the participants are normal and non-psychiatric according to the scores of the Beck's Depression Inventory and GHQ-28 tests. All participants are examined by a resident physician before the experiments (to check their blood pressure and inform them about the experimental environment) and fill the consent form for the participation in the experiments.

Retrieving positive autobiographical memories is the most applicable strategy used for upregulating positive emotion (Linhartová et al., 2019). The experimental paradigm in this study is similar to the one published in (Young et al., 2014; Vadim Zotev, Misaki, Phillips, Wong, & Bodurka, 2018; Vadim Zotev, Phillips, Yuan, Misaki, & Bodurka, 2014; Vadim Zotev et al., 2016) but with minor differences and is based on retrieving autobiographical happy memories. Before the experiment, each participant is interviewed and asked to write several positive autobiographical memories. The experiment contains 10 runs and each participant has 10 blocks of emotion regulation using various autobiographical memory retrievals. Each run contains three blocks including rest, view, and upregulation. The difference between our paradigm (to increase the effectiveness of autobiographical memories) and similar paradigms (Young et al., 2014; Vadim Zotev et al., 2014, 2016) is that we select pictures based on what subjects announce during interviews and present them during the experiments to remind the autobiographical memories related to these pictures in each run. The duration of the rest block is 20s, the view block is 40s, and the upregulation block is 60s. During the rest block, no image is shown, and only the message "please rest" is displayed on the screen to ask the participants to relax with no specific tasks. In the view block, two related pictures were presented for 40s and the participant is asked to see them without thinking about them or remembering anything. In the upregulation block, two images similar to the view block are presented and the participant tries to increase the height of the bar of neurofeedback based on the brain activity. The selected images for View and Upregulation blocks have similar arousal and valence (without any significant difference) according to the rates given after the experiment by each participant to individual images of positive autobiographical memories.

Before the experiment, we explain the paradigm to each participant by visual presentation of a sample run of the task and asking them to try to increase the height of the bar of the neurofeedback during the upregulation block by remembering the positive autobiographical memories cued by the depicted images collected based on what each participant announces during the interview (without showing the individual pictures). In each step, the bar is blue if the participant succeeds to increase or maintain the brain activity represented by neurofeedback and red if he does not succeed.

The neurofeedback in the experimental group is based on the approach–withdrawal hypothesis (Davidson, 1998), which is defined as the difference between the EEG power in the right and left hemispheres in the alpha frequency band and in the windows with a length of 2s, which is updated every 1s with 50% overlap between the consecutive windows (the height of the neurofeedback bar at each time is the average of the EEG asymmetry alpha power in the current and two previous windows). For the control group, the sham feedback is a randomly generated signal for each upregulation block as proposed in (Vadim Zotev et al., 2018). Neurofeedback is presented only in the upregulation block. The whole neurofeedback protocol for one run is depicted in Figure 1.

### 2.2 Data Acquisition

The MRI data are acquired using a 3.0 Tesla Siemens Prisma MRI Scanner located in the National Brain Mapping Lab (NBML), Tehran, Iran. Functional MRI are acquired using a T2*-weighted gradient-echo, echo-planar (EPI) pulse sequence (TR = 2000 msec, TE = 30 msec, matrix size = $64\times 64\times 30$, and voxel size = $3.8\times3.8\times4$ mm). During the 10 runs of the experiment session, 650 volume images are acquired. Structural images are acquired using a gradient-echo, T1-weighted MPRAGE pulse sequence (TI = 1100 msec, TR = 1810 msec, TE = 3.47 msec, and voxel size = $1\times1\times1$ mm).

The EEG data are recorded simultaneously with fMRI using the Brain Products EEG system. The EEG cap has 63 electrodes according to the 10-20 system and one ECG electrode. Electrode impedances are maintained below 5K Ohms by cleaning the skin and removing dirt using alcohol and injecting suitable gel. The EEG signal is recorded at 5K samples/sec. For quality control of the EEG data and neurofeedback, several minutes of recording are performed in the MRI control room just before the subject goes in the MRI scanner. The task is presented by Psych toolbox program through a coil mounted display, which allows the subjects to see the stimulus and different blocks of the experiment.

### 2.3 Real-time Data Processing

Due to practical limitations, neurofeedback is provided only based on the EEG signal. For the EEG neurofeedback, the RecView software is applied to remove MRI and ballistocardiogram artifacts from the EEG data in real time, using the moving average method (P. J. Allen, Josephs, & Turner, 2000; P. J. Allen, Polizzi, Krakow, Fish, & Lemieux, 1998). As the mean head displacement obtained in offline analysis is 0.41 mm, the result of the moving average template subtraction method will not differ significantly from the counterpart methods (Moosmann et al., 2009; Niazy, Beckmann, Iannetti, Brady, & Smith, 2005). In addition, since the head movement and ballistocadiogram artifacts induce large peaks in the EEG signal, the remaining large peaks are excluded. Removing the ballistocardiogram residual from the EEG signal is very difficult. However, since the mean head displacement is 0.41 mm and the frequency range of this artifact is less than 7 Hz, our feedback which is in the alpha frequency band will not be affected by the artifact.

The denoised data is down-sampled to 250 samples/sec. Then, powers of channels F3 and F4 are calculated every 1 sec using a 2 sec moving window. The relative EEG power asymmetry for F4 and F3 with respect to the baseline (by averaging the asymmetry values in the previous view block) is calculated and presented as a bar during upregulation blocks. For a high quality neurofeedback, the following steps are done. First, the power spectrum of the data recorded inside and outside of the MRI scanner is compared to evaluate quality of the denoised data. Secondly, the spectrum of neurofeedback and its values obtained by RecView are compared with the results of offline analysis to evaluate the neurofeedback quality. The results confirm the quality of the offline removal of the EEG artifacts and the reliability of the neurofeedback. They also prove that the neurofeedback provided to each participant is based on the actual brain activity, not the artifacts.

### 2.4 EEG Data Analysis

An offline analysis of the EEG data, acquired simultaneously with fMRI, is performed using FMRIB plug-in as a Matlab toolbox (Christov, 2004; K. H. Kim, Yoon, & Park, 2004; Niazy et al., 2005). This software removes the fMRI gradient artifacts, detects the QRS complexes from an ECG channel, and removes the pulse (ballistocardiographic/BCG) artifacts from the EEG signal. After removing the MRI and BCG artifacts, the EEG data are down-sampled to 250 samples/sec and low-pass filtered at 100 Hz. Next, the fMRI slice selection frequency and its harmonics are removed by bandpass filtering. Then, ICA is applied over the entire EEG data after excluding the noisy and motion-affected intervals. Next, independent components (ICs) corresponding to the artifacts, e.g., eye blinking, head movement, and cardioballistic or BCG residual, are identified and removed. This is done based on the time course spectral density (head movement in the range of 0.5-4.5 Hz, cardioballistic motion in the range of 2-7 Hz, eye blinking in the range of 0.5-3 Hz), topographic map (bipolar topography for BCG) (Mayeli, Zotev, Refai, & Bodurka, 2016), and kurtosis (rapid and random head movements have high kurtosis values) (Mognon, Jovicich, Bruzzone, & Buiatti, 2011; Wong et al., 2016; V Zotev, Yuan, Phillips, & Bodurka, 2012). Then, the average EEG power spectrum is calculated for each of the experimental blocks (Upregulation, View, and Rest). A moving window with a length of 2 sec and 50% interval overlap is applied on the EEG data to calculate the EEG asymmetry of channels F3 and F4 in the alpha band for each block.

### 2.5 fMRI Data Analysis

An offline analysis of the whole brain fMRI data is performed in FSL. Pre-processing of a single-subject fMRI data includes slice-timing correction, motion correction, temporal high pass filtering, and spatially smoothing using an 8 mm full-width at half-maximum Gaussian kernel. The standard GLM analysis is then applied to the fMRI time series. Three regressors for Upregulation, View, and Rest are convolved with the hemodynamic response function and six motion confounds are included in the GLM model. Finally, the whole brain is thresholded at p-value = 0.01 for voxels and for cluster correction at p-value = 0.01 in the cluster-level correction algorithm, which corrects for the multiple comparisons using the Gaussian Random Field (GRF) model (Fsl, 2006; Smith et al., 2004; Woolrich et al., 2009). To determine the active regions in the upregulation blocks, the contrasts of the “Upregulation versus View” and the “Upregulation versus Rest” are calculated. A global measure of the signal change in different activated regions of the preprocessed fMRI data registered to the Montreal Neurological Institute (MNI) atlas are calculated using an anatomical mask. For this purpose, the mean signal changes of the activated voxels in every active region for Upregulation versus View and Rest are calculated in the 118 anatomical masks extracted from the "WFU_PickAtlas" and FSL (Desikan et al., 2006;

Gorgolewski et al., 2015; Maldjian, Laurienti, & Burdette, 2004; Maldjian, Laurienti, Kraft, & Burdette, 2003; Tzourio-Mazoyer et al., 2002).

For further evaluation, the Pearson's correlation is used to estimate and compare the functional connectivity in the different blocks of the experiment. To this end, the anatomical masks of various brain regions are used to extract the mean BOLD signals of the activated voxels in the regions and calculate the functional connectivity between the regions. The results are used to reveal the neurobiological substrate of the neurofeedback protocol in the brain.

### 2.6 Psychometric Testing

To measure changes in the mood state, each participant completes Persian version of Beck Anxiety Inventory (BAI), State-Trait Anxiety Inventory (STAI), Positive-Negative affect scale (PANAS), and short Persian version of the Profile of Mood States (POMS) before and after the neurofeedback test. The Beck Anxiety Inventory (BAI) contains 21 questions and measures the severity of anxiety (Beck, Epstein, Brown, & Steer, 1988). The State-Trait Anxiety Inventory (STAI) includes 20 items for the state anxiety and 20 items for the trait anxiety (Azizi, Mohammadkhani, Foroughi, Lotfi, & Bahramkhani, 2013). The POMS includes a self-rating of six different aspects of the mood, e.g., Tension or Anxiety, Anger or Hostility, Vigor or Activity, Fatigue or Inertia, Depression or Dejection, and Confusion or Bewilderment, with a five-points scale rating from "not at all" to "extremely" (Spielberger, 1972). The PANAS also contains a self-rating of the positive and negative affects with a five-points scale rating from "not at all" to " very much" (Watson, Clark, & Tellegen, 1988).

## 3. Results

### 3.1 EEG Results

The EEG power in the alpha band is used for neurofeedback. The average change of the alpha power asymmetry in the Upregulation, View and Rest blocks for the experimental and control groups are illustrated in Figure 2. Since the feedback for the experimental group is based on the powers of channels F4 and F3, and the subjects try to increase LnP(F4)-LnP(F3) in the alpha frequency band versus the view period, we expect the mean power in the upregulation block to be higher than the rest and view blocks for the experimental group as illustrated in Figure 2(a). The changes in the EEG asymmetry for the Upregulation versus View/Rest periods are significant for the experimental group (paired t-test; $df = 179$; $\text{p-value}_{\text{Upregulation-View}} = 5.1 \times 10^{-5}$, $\text{p-value}_{\text{Upregulation-Rest}} = 3.8 \times 10^{-6}$) but not for the control group. Also, the changes in the EEG frontal asymmetry for the Upregulation versus View and Rest in the experimental group are significantly different from those in the control group (two sample t-test; df = 318; p-value $_{\text{Upregulation-View (Experimental group-Control group)}}$ = $1.7 \times 10^{-4}$, p-value $_{\text{Upregulation-Rest (Experimental group-Control group)}}$ = $3.9 \times 10^{-3}$).

Figures 2(b) and 2(c) show the effectiveness of the neurofeedback protocol as the difference between the frontal asymmetry of upregulation and view/rest in the experimental group is higher than in the control group. Consistent with the approach–withdrawal model that states during happiness, the difference between the power of the right and left hemispheres increases, the difference between the EEG power of channels F4 and F3 in the upregulation blocks is higher than the view and rest blocks.

To evaluate the effect of heart rate variability (HRV) on the emotion regulation, HRV was extracted for all participants in three blocks of the experiment, using their ECG channel recorded during the experiment. The results show that HRV in the upregulation block is significantly different from those of the view and rest blocks in only 3 of the 32 participants. This indicates that the emotion regulation is not associated with HRV.

### 3.2 fMRI Results

Activation detection in the whole brain using GLM with different contrasts of Upregulation versus View and Rest illustrate several regions for the experimental group during emotion regulation. As mentioned earlier, to have a global evaluation, we use an anatomical mask for each region to compare the mean signal change of the activated regions between the Upregulation and Rest/View blocks. To remove the effect of hemodynamic response of the View block on the Upregulation block, the first 4 sec (two samples) of the BOLD signal of each Upregulation block is removed for signal change calculation and functional connectivity analysis. The results of significant signal changes for the Upregulation versus View and Rest (paired t-test for every one of the 118 ROIs used in this study; df = 179; FDR-corrected for multiple comparison; q = 0.01) and the corresponding t-value and Cohen's D effect size (Cohen, 1988) for the experimental group are listed in Table 1. They illustrate several brain regions with large signal change and effect size for emotion regulation, which are consistent with the literature (listed in the last column of Table 1) (Bado et al., 2014; Burianova, McIntosh, & Grady, 2010; Ino, Nakai, Azuma, Kimura, & Fukuyama, 2011; S. J. Johnston, Boehm, Healy, Goebel, & Linden, 2010; S. Johnston et al., 2011; S. H. Kim & Hamann, 2007; Lempert, Speer, Delgado, & Phelps, 2017; Li et al., 2016; Pelletier et al., 2003; Vadim Zotev et al., 2016, 2014). The changes in the BOLD signal of brain regions in Table 1 for the Upregulation versus View and Rest in the experimental group are significantly different from those of the control group (two sample t-test; df = 318; FDR-corrected for multiple comparison; q = 0.05).

Multiple brain regions especially in the sub-cortical regions play key roles in emotion regulation. The percentage of the signal change in the brain regions listed in Table 1 is higher than those of the previous studies. This may be due to the fact that the proposed emotion regulation paradigm (neurofeedback) is based on the recall of the positive autobiographical memory using the stimulus pictures related to the issues announced by the participants to induce happiness. Figure 3 presents the activation map for the Upregulation versus View in the experimental group.

Figure 4 provides the box plot of the signal changes in the most important regions for the rest, view, and upregulation states for all subjects in the experimental groups. In Figure 4, the signal change of each experimental group participant for each ROI is computed by subtracting a global minimum from the average intensity of that ROI and normalizing the result by the "difference between the global minimum and the global maximum." Figure 4 confirms the effectiveness of the proposed neurofeedback protocol during upregulation. It shows higher signal changes for the upregulation relative to the view and rest for the experimental group.

The results of the experimental group in Figure 4 reveal the effect of neurofeedback during emotion regulation. To have a better understanding of the neurobiological substrate of emotion regulation by the EEG neurofeedback and separate the effect of the neurofeedback and the recall of the autobiographical memories on upregulation, the functional connectivity of the thirty-eight brain regions (including left/right amygdala, thalamus, insula, dorsomedial prefrontal cortex, caudate, cuneus, hippocampus, posterior cingulate cortex, orbitofrontal cortex, middle temporal gyrus, lingual gyrus, ventral striatum, dorsolateral prefrontal cortex, ventrolateral prefrontal cortex, superior parietal, inferior parietal, supramarginal, postcentral and anterior cingulate cortex) are calculated during each block of the paradigm. Then, the significant connections between the Upregulation and View (paired t-test for $38 \times (38-1)/2 = 703$ different connections; df = 179; FDR-corrected for multiple comparison; q = 0.05) for the experimental group are obtained. Functional connectivity is estimated by cross correlation. Next, to reveal the effect of the neurofeedback, the significant links extracted from the experimental group are compared with those of the control group (two sample t-test; df = 318; FDR-corrected for multiple comparison; q=0.05). Connections significantly different between the two groups are due to the neurofeedback. Each illustrated edge in Figure 5 shows a connection with a significant change between the upregulation and view blocks (and between the experimental and control groups) as a result of the neurofeedback. A ticker line between the regions corresponds to a higher correlation value.

Figure 5 illustrates the functional connectivity of the brain regions. Functional connectivity between several emotion-related regions increases in the upregulation block, e.g., left amygdala and left thalamus. The roles of regions such as amygdale, insula, thalamus, left anterior cingulated cortex, hippocampus, orbitofrontal cortex, and ventrolateral prefrontal cortex in emotion regulation and recalling autobiographical happy memories are described in the Discussion Section.

To demonstrate the distinction between the functional connectivity of the upregulation and view blocks, the functional connectivity distribution of the significant connections in the experimental group is shown in Figure 6. The distributions of the upregulation and view blocks show higher functional connectivity in the upregulation blocks and therefore more synchronization among the brain regions during emotion regulation.

### 3.3 Mood Assessment

The emotional state test results obtained before and after the neurofeedback for the experimental group are reported in Table 2. The results confirm the effectiveness of the neurofeedback. The average scores for PANAS does not change significantly due to the neurofeedback but for the positive and negative mood states of PANAS, POMS, and Total Mood distribution (TMD), the changes are significant (paired t-test; df=17; $p_{\text{positive mood states of PANAS}} = 4.5 \times 10^{-4}$, $p_{\text{negative mood states of PANAS}} = 2 \times 10^{-3}$, $p_{\text{POMS}} = 1.1 \times 10^{-3}$, $p_{\text{TMD}} = 2.1 \times 10^{-4}$). This demonstrates that neurofeedback is effective in increasing the positive mood state and decreasing the negative mood state through recalling positive autobiographical memories. Changing the mean value of TMD from 7.5 to -4.7 may be interpreted as an increase in the positive mood and a decrease in the negative mood as a result of the neurofeedback.

The emotional state tests of before and after the neurofeedback for the control group are reported in Table 3. The scores of PANAS negative mood states show a change towards significance from before to after the neurofeedback (paired t-test; df=13; p-value = 0.064) but for the positive mood states, they do not change significantly. This demonstrates that the recalling of the positive autobiographical memory even with a sham neurofeedback is effective in decreasing (increasing) the negative (positive) mood state of the control group. The scores of POMS and Total Mood distribution (TMD) do not change significantly from before to after the neurofeedback. Changing the mean value of TMD from 6.6 to 3.6 for the control group may be interpreted as increasing the positive mood states and decreasing the negative mood states by the recalling of the autobiographical memories even with a sham neurofeedback. There are significant differences between the changes of the positive mood states of PANAS and TMD due to the neurofeedback in the experimental and control groups (two sample t-test; df = 30; $p_{\text{positive mood states of PANAS (Experimental group-Control group)}} = 0.02$ and $p_{\text{TMD (Experimental group-Control group)}} = 0.004$).

## 4. Discussion

In this study, the participants were able to upregulate positive emotion using the EEG neurofeedback based on the frontal asymmetry in the alpha frequency band and a happiness induced task through retrieving their positive autobiographical memories. Comparison of the results of the experimental and control groups reveals increased activity of the prefrontal, insular, and limbic regions and increased functional connectivity in/between the prefrontal, limbic, and insular regions as a result of the EEG neurofeedback in the experimental group. The psychometric tests confirm an increased positive emotion and a decreased negative emotion as a result of the neurofeedback.

Before and after the neurofeedback experiment, the Persian version of the Beck Anxiety Inventory (BAI) and State-Trait Anxiety Inventory (STAI) were completed by each participant to measure and evaluate the anxiety that might be induced by the fMRI scan. The results showed that the anxiety scores of all participants before and after neurofeedback experiment were less than 16 and 33 for BAI and State Anxiety Inventory of STAI (in the range of minimal and mild anxiety), and after the neurofeedback test, the level of anxiety slightly decreased. Therefore, not only the anxiety did not increase during the experiment but also decreased as a result of the neurofeedback and emotion regulation. The mean ± standard deviation of the Beck Anxiety Inventory (BAI) and State Anxiety Inventory for “before neurofeedback” were 8 ± 4.4 and 27 ± 2.5 and for “after neurofeedback” were 6.6 ± 4.7 and 26.3 ± 2.8.

As mentioned in the previous studies, several brain areas including the subcortical and limbic regions are involved in emotion regulation. Consistent with the previous studies, during emotion regulation, amygdala, insula, anterior cingulate cortex, cuneus, caudate, orbitofrontal cortex, ventrolateral prefrontal cortex, ventral striatum, and temporal gyrus are activated (Table 1). As shown in Figure 5, the functional connectivity between several emotional related regions changes during the upregulation blocks relative to the view blocks.

The limbic system includes regions such as hypothalamus, thalamus, amygdala, and hippocampus with key roles in emotion regulation. Amygdala has a key role in emotion regulation and emotion generation and with hippocampus and prefrontal cortex plays an important role in the retrieval of both positive and negative autobiographies (Bracht et al., 2009). Amygdala connects with several emotion regulatory regions such as medial and lateral orbitofrontal cortices, anterior cingulate cortex (ACC), and DLPFC through ventral and dorsal pathways (Ochsner & Gross, 2008). The activity of amygdala increases during various stimulus including the observation of the positive and negative pictures. Positive correlation between the BOLD signal of amygdala and the recall of the positive pictures is reported in a previous study (Phelps, 2004).

Thalamus is connected to several emotion regulatory areas including amygdala, insula, and hippocampus and executes several brain functions such as emotion, motivation, executive function, learning, and decision-making (Maddock, Garrett, & Buonocore, 2001). The involvement of thalamus in emotion regulation and autobiographical memories is demonstrated in a meta-analysis of the fMRI studies. Thalamus activates in a wide range of positive and negative emotional stimuli like happiness, sadness, and disgust (Zotev et al., 2018). Thalamus is considered as a relay center for most of the sensory information. The sensory information first stops in thalamus and then goes to destinations in the cortex. Therefore, a greater amount of processing in situations like happiness is associated with higher thalamus activity (Cerqueira et al., 2008).

Insula is a part of the emotion regulatory network and like previous studies of emotion regulation, the activity of the right and left insula increases during valence or cognitive tasks such as disgust and autobiographical happy memory (Pohl et al., 2013). Insula plays an important role in emotion processing and monitors internal emotional states like disgust, happiness, and sadness (Chen et al., 2009). Increased insula activity indicates its role in receiving emotional signals from various brain regions. Insula coordinates the brain networks through its functional connectivity with other brain regions like thalamus and caudate during different tasks.

Several studies showed increased activity of the left/right anterior cingulate cortex (ACC) when recalling happy memories (Bush et al., 2000). ACC is connected to several limbic regions and prefrontal and frontal cortices and its activity increases during cognitive reappraisal and emotion regulation and decreases during expressive suppression (Mitterschiffthaler, Fu, Dalton, Andrew, & Williams, 2007; Suardi, Sotgiu, Costa, Cauda, & Rusconi, 2016).

Increased activity of the lingual gyrus is reported in working memory related to visual information and memory retrieval and in visual imagery tasks (Burianova & Grady, 2007; Burianova et al., 2010; Gilboa, Winocur, Grady, Hevenor, & Moscovitch, 2004). Since the task used in our work includes visual information related to happy autobiographical memory, we expect an increased activity of the lingual gyrus.

Increased activity in the cuneus is reported in several tasks including cognitive functions, visual processing, visuo-spatial imagery, and episodic memory retrieval. In our study, we induced happy autobiographical memories using related images and found increased activity in the cuneus (Deak et al., 2017; Vrticka et al., 2013).

Orbitofrontal cortex (OFC) is a part of the frontal cortex and is a key region in emotion regulation and self-monitoring with reciprocal connections with the other related regions in emotion regulation such as amygdala and anterior cingulate cortex (Schutter & van Honk, 2006). Involvement of OFC in the retrieval of positive emotion is shown in the previous studies (Beer,

John, Scabini, & Knight, 2006). OFC acts as a filter or gate for the neural activity of the subcortical regions initiated by an emotional task and then monitors the emotional responses of the other brain regions (Kringelbach & Rolls, 2004).

As a part of the limbic system, ventral striatum is another key region in emotion regulation. Ventral activity is related to its role in reward processing (Marci et al., 2007). Therefore, increased activity of the ventral striatum during happiness or autobiographical happy memory can be interpreted as a rewarding process or state. Positive events impact the processing and encoding of information in the memory and engage the reward related regions such as ventral striatum and enhance recalling (Hare, Tottenham, Davidson, Glover, & Casey, 2005).

Ventrolateral prefrontal cortex is another region involved in emotion generation and regulation. Several researches have confirmed the increased activity and the role of the ventrolateral prefrontal cortex in recalling autobiographical memory (Svoboda et al., 2006). According to the proposed models for the ventrolateral prefrontal cortex, this region has functional connectivity with amygdala and is activated automatically with an emotional stimulus. In the proposed task, we induced autobiographical happy memory. Therefore, an increased activity in the ventrolateral prefrontal cortex was expected (He et al., 2018).

Dorsolateral prefrontal cortex (DLPFC) as a part of the prefrontal region associates with cognitive functions, working memory, decision making, and pleasant emotional stimulus. As discovered by previous studies, bilaterally increased activity in DLPFC is a result of positive stimulus (Herrington et al., 2005).

The increased activity of the parietal regions (superior parietal, inferior parietal, supramarginal and postcentral) is related to the role of these regions in attention deployment to the presented images of positive autobiographical memories and their involvement in the integration of the sensory and behavioral information (Aday, Rizer, & Carlson, 2017; Andersen, 1997; Bullier, 2001).

According to the emotion regulation model described in (Kohn et al., 2014), emotion regulation is modelled as a process with three steps. The first step is an appraisal of the stimulus. Subcortical regions such as amygdala, basal ganglia, and ventral striatum play a key role in emotion generation. In this step, the affective arousal is relayed and projected to VLPFC via subcortical regions (amygdala, basal ganglia, and ventral striatum). The next step is “detecting and start of regulation process” and the need for regulation is signaled by VLPFC and insula. The final step is regulation and change of the emotional state and moving to a new state. Thalamus directs the sensory information to different cortical regions. Therefore, increasing the functional connectivity between amygdala and thalamus and between ventral striatum and thalamus are the results of emotion generation by amygdala and ventral striatum and transmission of the somatosensory information to the cortical regions by thalamus in emotion regulation. The increased functional connectivity between thalamus and insula can be interpreted as the role of insula in the second step of emotion regulation for sending the need for regulation to the cortical regions and its involvement in the third step of emotion regulation and the role of thalamus as the relay center for sending sensory information to cortical regions.

The increased functional connectivity between thalamus and VLPFC may be interpreted similar to those of thalamus and insula. According to the proposed model in (Kohn et al., 2014), VLPFC and insula have similar roles in emotion regulation. OFC receives input and sensory information from various brain regions such as amygdala, hypothalamus, and the insular cortex. Therefore, the rise of connectivity between OFC and thalamus originates from relaying sensory information by thalamus and receiving it by OFC. There are indirect connections between the subcortical and prefrontal regions involved in emotion regulation. The functional connectivity between OFC and VLPFC, DMPFC and ventral striatum, DMPFC and amygdala, DMPFC and insula, thalamus and DMPFC are supported by extensive and reciprocal anatomical connections in the prefrontal cortex, between the prefrontal cortex and the limbic/paralimbic regions, and through indirect connection according to the steps of the emotion regulation models.

The increased functional connectivity between thalamus and postcentral (as a part of parietal lobe) is related to the role of parietal to receive and process sensory information (like visual information) from thalamus as a relay center for sending sensory information.

Frontal regions especially OFC has important functional connectivity with sub-cortical regions including thalamus and caudate in major depression disorder and associates with emotional behaviors like decision-making. Lower connectivity correlates with more suicidal ideation in the major depression disorder (Cheng et al., 2016).

The functional connectivity results of this study is consistent with the previous emotion regulation studies and models. There are some new connections identified in this study which are not mentioned in the previous studies. The new connections are between thalamus and dorsomedial prefrontal cortex, thalamus and insula, thalamus and orbitofrontal cortex, thalamus and ventrolateral prefrontal cortex, thalamus and postcentral (parietal), dorsomedial prefrontal cortex and ventral striatum, and ventrolateral prefrontal cortex and orbitofrontal cortex. In Table 4, the connectivity identified in this study and those of the previous studies are presented. The emotion regulation paradigms and strategies for some of the previous researches reported in Table 4 are different from this study. We reported them because of the concept of emotion regulation used in those studies and also the existence of some brain circuits activated regardless of the type of stimulus. We also wanted to illustrate that some of the connections identified in this study were not reported in the previous emotion regulation studies (even with different paradigms).

## 5. Conclusion

This study demonstrates the effectiveness of the EEG neurofeedback in changing EEG and fMRI signals. It also shows changes in the functional connectivity of the brain regions involved in emotion regulation, especially the prefrontal, limbic, and insular regions. The EEG neurofeedback using the proposed happy autobiographical memory paradigm changes the fMRI BOLD signal of various brain regions and the EEG frontal asymmetry more than those observed in the previous studies, even when they used both modalities for neurofeedback (S. J. Johnston et al., 2010; Li et al., 2016; Young et al., 2014; Vadim Zotev et al., 2014, 2016). Comparison of the psychometric test results, obtained before and after the neurofeedback experiment, confirms the effectiveness of the neurofeedback paradigm in changing the negative and positive mood states of the participants. In addition, comparison of the results of the experimental and control groups reveals the effectiveness of the neurofeedback on the emotion upregulation with significant changes in the experimental group.

Through simultaneous recording of fMRI with the EEG neurofeedback, the functional connectivity of various brain regions, especially between the frontal, parietal, limbic, and insular regions, reveals the involved network as a neurobiological substrate of emotion regulation by neurofeedback, and separates the effect of the neurofeedback and the recall of the autobiographical memories. We detected some new connectivity links of this network. They are reported in Table 4 and exist between thalamus and dorsomedial prefrontal cortex, thalamus and insula, thalamus and orbitofrontal cortex, thalamus and ventrolateral prefrontal cortex, thalamus and postcentral (parietal), dorsomedial prefrontal cortex and ventral striatum, ventrolateral prefrontal cortex and orbitofrontal cortex. They are justified through emotion regulation models (Kohn et al., 2014). The results demonstrated that Upregulation though EEG neurofeedback provided changes in activity and interaction of involved regions in emotion regulation like those observed in the previous emotion regulation and neurofeedback studies with high signal changes and effect size. The increased functional connectivity between prefrontal/frontal and other brain regions may provide valuable information, particularly in the EEG neurofeedback studies, where the prefrontal activity can be measured by EEG and modulation/activity of other regions can be predicted. The results may suggest the use of the proposed paradigm for the treatment of the mental disorders in a small number of treatment sessions because of the larger changes and effect size in the EEG and fMRI signals and the psychometric assessments.

The proposed connectivity analysis may be used in the neurofeedback studies in addition to the powers of EEG and the fMRI BOLD signals. The brain is a complex network and the function of each region affects the others. In most of the brain functions and mental disorders, multiple brain regions are involved. Therefore, using feedback based on the connectivity of the involved regions for changing the whole network for a specific paradigm or mental disorder may be more effective than the alternative methods and would be an avenue of further investigation (Linhartová et al., 2019; Sulzer et al., 2013).

## Acknowledgment

The authors would like to thank National Brain Mapping Lab (NBML), Tehran, Iran for assistance with the experiments and providing technical support.

## Author Contributions

All authors listed have made a substantial, direct and intellectual contribution to the work, and approved it for publication.

## Conflict of interest

The authors have no conflict of interests or financial disclosures to declare.

## Ethical approval

The research protocol is approved by the ethics committees of the Iran University of Medical Sciences, Tehran, Iran.

## Informed consent

Informed consent was obtained from all individual subjects included in the study.

## Data Availability Statement

The data that support the findings of this study are available on request from the corresponding author. The data are not publicly available due to privacy or ethical restrictions.

## References

Aday, J., Rizer, W., & Carlson, J. M. (2017). Neural Mechanisms of Emotions and Affect. In *Emotions and Affect in Human Factors and Human-Computer Interaction*. https://doi.org/10.1016/B978-0-12-801851-4.00002-1

Allen, J. J. B., & Reznik, S. J. (2015). Frontal EEG asymmetry as a promising marker of depression vulnerability: Summary and methodological considerations. *Current Opinion in Psychology*. https://doi.org/10.1016/j.copsyc.2014.12.017

Allen, P. J., Josephs, O., & Turner, R. (2000). A method for removing imaging artifact from continuous EEG recorded during functional MRI. *NeuroImage*. https://doi.org/10.1006/nimg.2000.0599

Allen, P. J., Polizzi, G., Krakow, K., Fish, D. R., & Lemieux, L. (1998). Identification of EEG events in the MR scanner: The problem of pulse artifact and a method for its subtraction. *NeuroImage*. https://doi.org/10.1006/nimg.1998.0361

Andersen, R. A. (1997). Multimodal integration for the representation of space in the posterior parietal cortex. In *Philosophical Transactions of the Royal Society B: Biological Sciences*. https://doi.org/10.1098/rstb.1997.0128

Azizi, A., Mohammadkhani, P., Foroughi, A. akbar, Lotfi, S., & Bahramkhani, M. (2013). The Validity and Reliability of the Iranian Version of the Self-Compassion Scale. *Practice in Clinical Psychology*.

Bado, P., Engel, A., de Oliveira-Souza, R., Bramati, I. E., Paiva, F. F., Basilio, R., … Moll, J. (2014). Functional dissociation of ventral frontal and dorsomedial default mode network components during resting state and emotional autobiographical recall. *Human Brain Mapping*. https://doi.org/10.1002/hbm.22403

Banks, S. J., Eddy, K. T., Angstadt, M., Nathan, P. J., & Luan Phan, K. (2007). Amygdala-frontal connectivity during emotion regulation. *Social Cognitive and Affective Neuroscience*. https://doi.org/10.1093/scan/nsm029

Bartholdy, S., Musiat, P., Campbell, I. C., & Schmidt, U. (2013). The potential of neurofeedback in the treatment of eating disorders: a review of the literature. *European Eating Disorders Review : The Journal of the Eating Disorders Association*. https://doi.org/10.1002/erv.2250

Beck, A. T., Epstein, N., Brown, G., & Steer, R. A. (1988). An Inventory for Measuring Clinical Anxiety: Psychometric Properties. *Journal of Consulting and Clinical Psychology*. https://doi.org/10.1037/0022-006X.56.6.893

Beer, J. S., John, O. P., Scabini, D., & Knight, R. T. (2006). Orbitofrontal cortex and social behavior: Integrating self-monitoring and emotion-cognition interactions. *Journal of Cognitive Neuroscience*. https://doi.org/10.1162/jocn.2006.18.6.871

Bolea, A. S. (2010). Neurofeedback treatment of chronic inpatient schizophrenia. *Journal of Neurotherapy*. https://doi.org/10.1080/10874200903543971

Bracht, T., Tüscher, O., Schnell, S., Kreher, B., Rüsch, N., Glauche, V., … Saur, D. (2009). Extraction of prefronto-amygdalar pathways by combining probability maps. *Psychiatry Research - Neuroimaging*. https://doi.org/10.1016/j.pscychresns.2009.05.001

Bullier, J. (2001). Integrated model of visual processing. *Brain Research Reviews*. https://doi.org/10.1016/S0165-0173(01)00085-6

Burianova, H., & Grady, C. L. (2007). Common and unique neural activations in autobiographical, episodic, and semantic retrieval. *Journal of Cognitive Neuroscience*. https://doi.org/10.1162/jocn.2007.19.9.1520

Burianova, H., McIntosh, A. R., & Grady, C. L. (2010). A common functional brain network for autobiographical, episodic, and semantic memory retrieval. *NeuroImage*. https://doi.org/10.1016/j.neuroimage.2009.08.066

Bush, G., Luu, P., & Posner, M. (2000). Cognitive and emotional influences in anterior cingulate cortex. *Trends in Cognitive Sciences*. https://doi.org/10.1016/S1364-6613(00)01483-2

Cavazza, M., Aranyi, G., Charles, F., Porteous, J., Gilroy, S., Klovatch, I., … Raz, G. (2014). Towards Empathic Neurofeedback for Interactive Storytelling. *Proceedings of the 5th Workshop on Computational Models of Narrative (CMN'14)*. https://doi.org/10.4230/OASIcs.CMN.2014.42

Cerqueira, C. T., Almeida, J. R. C., Gorenstein, C., Gentil, V., Leite, C. C., Sato, J. R., … Busatto, G. F. (2008). Engagement of multifocal neural circuits during recall of autobiographical happy events. *Brazilian Journal of Medical and Biological Research*. https://doi.org/10.1590/S0100-879X2008001200006

Chen, Y. H., Dammers, J., Boers, F., Leiberg, S., Edgar, J. C., Roberts, T. P. L., & Mathiak, K. (2009). The temporal dynamics of insula activity to disgust and happy facial expressions: A magnetoencephalography study. *NeuroImage*. https://doi.org/10.1016/j.neuroimage.2009.04.093

Cheng, W., Rolls, E. T., Qiu, J., Liu, W., Tang, Y., Huang, C. C., … Feng, J. (2016). Medial reward and lateral non-reward orbitofrontal cortex circuits change in opposite directions in depression. *Brain*. https://doi.org/10.1093/brain/aww255

Christov, I. I. (2004). Real time electrocardiogram QRS detection using combined adaptive threshold. *BioMedical Engineering Online*. https://doi.org/10.1186/1475-925X-3-28

Coan, J. A., & Allen, J. J. B. (2004). Frontal EEG asymmetry as a moderator and mediator of emotion. *Biological Psychology*. https://doi.org/10.1016/j.biopsycho.2004.03.002

Coben, R., Linden, M., & Myers, T. E. (2010). Neurofeedback for autistic spectrum disorder:

A review of the literature. *Applied Psychophysiology Biofeedback*. https://doi.org/10.1007/s10484-009-9117-y

Cohen, J. (1988). *Statistical power analysis for the behavioral sciences (2nd ed.). Hillsdale, NJ: Lawrence Earlbaum Associates. Lawrence Earlbaum Associates*.

Cohen Kadosh, K., Luo, Q., de Burca, C., Sokunbi, M. O., Feng, J., Linden, D. E. J., & Lau, J. Y. F. (2016). Using real-time fMRI to influence effective connectivity in the developing emotion regulation network. *NeuroImage*. https://doi.org/10.1016/j.neuroimage.2015.09.070

Davidson, R. J., Ekman, P., Saron, C. D., Senulis, J. A., & Friesen, W. V. (1990). Approach-Withdrawal and Cerebral Asymmetry: Emotional Expression and Brain Physiology I. *Journal of Personality and Social Psychology*. https://doi.org/10.1037/0022-3514.58.2.330

Deak, A., Bodrogi, B., Biro, B., Perlaki, G., Orsi, G., & Bereczkei, T. (2017). Machiavellian emotion regulation in a cognitive reappraisal task: An fMRI study. *Cognitive, Affective and Behavioral Neuroscience*. https://doi.org/10.3758/s13415-016-0495-3

DeBettencourt, M. T., Cohen, J. D., Lee, R. F., Norman, K. A., & Turk-Browne, N. B. (2015). Closed-loop training of attention with real-time brain imaging. *Nature Neuroscience*. https://doi.org/10.1038/nn.3940

Dennis, T. A., & Solomon, B. (2010). Frontal EEG and emotion regulation: Electrocortical activity in response to emotional film clips is associated with reduced mood induction and attention interference effects. *Biological Psychology*. https://doi.org/10.1016/j.biopsycho.2010.09.008

Desikan, R. S., Ségonne, F., Fischl, B., Quinn, B. T., Dickerson, B. C., Blacker, D., … Killiany, R. J. (2006). An automated labeling system for subdividing the human cerebral cortex on MRI scans into gyral based regions of interest. *NeuroImage*. https://doi.org/10.1016/j.neuroimage.2006.01.021

Fsl. (2006). FMRIB Software Library.

Gapen, M., van der Kolk, B. A., Hamlin, E., Hirshberg, L., Suvak, M., & Spinazzola, J. (2016). A Pilot Study of Neurofeedback for Chronic PTSD. *Applied Psychophysiology Biofeedback*. https://doi.org/10.1007/s10484-015-9326-5

Gilboa, A., Winocur, G., Grady, C. L., Hevenor, S. J., & Moscovitch, M. (2004). Remembering our past: Functional neuroanatomy of recollection of recent and very remote personal events. *Cerebral Cortex*. https://doi.org/10.1093/cercor/bhh082

Gorgolewski, K. J., Varoquaux, G., Rivera, G., Schwarz, Y., Ghosh, S. S., Maumet, C., …

Margulies, D. S. (2015). NeuroVault.org: a web-based repository for collecting and sharing unthresholded statistical maps of the human brain. *Frontiers in Neuroinformatics*. https://doi.org/10.3389/fninf.2015.00008

Gross, J. J. (1998). The emerging field of emotion regulation: An integrative review. *Review of General Psychology*. https://doi.org/10.1037/1089-2680.2.3.271

Hammer, B. U., Colbert, A. P., Brown, K. A., & Ilioi, E. C. (2011). Neurofeedback for insomnia: A pilot study of Z-score SMR and individualized protocols. *Applied Psychophysiology Biofeedback*. https://doi.org/10.1007/s10484-011-9165-y

Hammond, D. C. (2003). QEEG-Guided Neurofeedback in the Treatment of Obsessive Compulsive Disorder. *Journal of Neurotherapy*. https://doi.org/10.1300/J184v07n02_03

Hare, T. A., Tottenham, N., Davidson, M. C., Glover, G. H., & Casey, B. J. (2005). Contributions of amygdala and striatal activity in emotion regulation. *Biological Psychiatry*. https://doi.org/10.1016/j.biopsych.2004.12.038

HE, W., BU, H., GAO, H., TONG, L., WANG, L., LI, Z., & YAN, B. (2017). Altered Amygdala Information Flow during rt-fMRI Neurofeedback Training of Emotion Regulation. *DEStech Transactions on Computer Science and Engineering*. https://doi.org/10.12783/dtcse/smce2017/12436

He, Z., Lin, Y., Xia, L., Liu, Z., Zhang, D., & Elliott, R. (2018). Critical role of the right VLPFC in emotional regulation of social exclusion: a tDCS study. *Social Cognitive and Affective Neuroscience*. https://doi.org/10.1093/scan/nsy026

Herrington, J. D., Mohanty, A., Koven, N. S., Fisher, J. E., Stewart, J. L., Banich, M. T., … Heller, W. (2005). Emotion-modulated performance and activity in left dorsolateral prefrontal cortex. *Emotion*. https://doi.org/10.1037/1528-3542.5.2.200

Herwig, U., Lutz, J., Scherpiet, S., Scheerer, H., Kohlberg, J., Opialla, S., … Brühl, A. B. (2019). Training emotion regulation through real-time fMRI neurofeedback of amygdala activity. *NeuroImage*. https://doi.org/10.1016/j.neuroimage.2018.09.068

Ino, T., Nakai, R., Azuma, T., Kimura, T., & Fukuyama, H. (2011). Brain activation during autobiographical memory retrieval with special reference to default mode network. *The Open Neuroimaging Journal*. https://doi.org/10.2174/1874440001105010014

Johnston, S. J., Boehm, S. G., Healy, D., Goebel, R., & Linden, D. E. J. (2010). Neurofeedback: A promising tool for the self-regulation of emotion networks. *NeuroImage*. https://doi.org/10.1016/j.neuroimage.2009.07.056

Johnston, S., Linden, D. E. J., Healy, D., Goebel, R., Habes, I., & Boehm, S. G. (2011). Upregulation of emotion areas through neurofeedback with a focus on positive mood.

*Cognitive, Affective and Behavioral Neuroscience*. https://doi.org/10.3758/s13415-010-0010-1

Kaur, C., & Singh, P. (2017). Toward EEG spectral analysis of tomographic neurofeedback for depression. In *Advances in Intelligent Systems and Computing*. https://doi.org/10.1007/978-981-10-1708-7_10

Kim, D. Y., Yoo, S. S., Tegethoff, M., Meinlschmidt, G., & Lee, J. H. (2015). The inclusion of functional connectivity information into fmri-based neurofeedback improves its efficacy in the reduction of cigarette cravings. *Journal of Cognitive Neuroscience*. https://doi.org/10.1162/jocn_a_00802

Kim, K. H., Yoon, H. W., & Park, H. W. (2004). Improved ballistocardiac artifact removal from the electroencephalogram recorded in fMRI. *Journal of Neuroscience Methods*. https://doi.org/10.1016/j.jneumeth.2003.12.016

Kim, S. H., & Hamann, S. (2007). Neural Correlates of Positive and Negative Emotion Regulation. *Journal of Cognitive Neuroscience*. https://doi.org/10.1162/jocn.2007.19.5.776

Kinreich, S., Podlipsky, I., Intrator, N., & Hendler, T. (2012). Categorized EEG neurofeedback performance unveils simultaneous fMRI deep brain activation. In *Lecture Notes in Computer Science (including subseries Lecture Notes in Artificial Intelligence and Lecture Notes in Bioinformatics)*. https://doi.org/10.1007/978-3-642-34713-9_14

Kohn, N., Eickhoff, S. B., Scheller, M., Laird, A. R., Fox, P. T., & Habel, U. (2014). Neural network of cognitive emotion regulation - An ALE meta-analysis and MACM analysis. *NeuroImage*. https://doi.org/10.1016/j.neuroimage.2013.11.001

Koole, S. (2009). The psychology of emotion regulation: An integrative review. *Cognition and Emotion*. https://doi.org/10.1080/02699930802619031

Koush, Y., Meskaldji, D. E., Pichon, S., Rey, G., Rieger, S. W., Linden, D. E. J., … Scharnowski, F. (2017). Learning Control Over Emotion Networks Through Connectivity-Based Neurofeedback. *Cerebral Cortex (New York, N.Y. : 1991)*. https://doi.org/10.1093/cercor/bhv311

Koush, Y., Pichon, S., Eickhoff, S. B., Van De Ville, D., Vuilleumier, P., & Scharnowski, F. (2019). Brain networks for engaging oneself in positive-social emotion regulation. *NeuroImage*. https://doi.org/10.1016/j.neuroimage.2018.12.049

Koush, Y., Rosa, M. J., Robineau, F., Heinen, K., W. Rieger, S., Weiskopf, N., … Scharnowski, F. (2013). Connectivity-based neurofeedback: Dynamic causal modeling for real-time fMRI. *NeuroImage*. https://doi.org/10.1016/j.neuroimage.2013.05.010

Kringelbach, M. L., & Rolls, E. T. (2004). The functional neuroanatomy of the human orbitofrontal cortex: Evidence from neuroimaging and neuropsychology. *Progress in Neurobiology*. https://doi.org/10.1016/j.pneurobio.2004.03.006

Kubik, A., & Biedroń, A. (2013). Neurofeedback therapy in patients with acute and chronic pain syndromes--literature review and own experience. *Przegląd Lekarski*.

Lackner, N., Unterrainer, H. F., Skliris, D., Wood, G., Wallner-Liebmann, S. J., Neuper, C., & Gruzelier, J. H. (2016). The Effectiveness of Visual Short-Time Neurofeedback on Brain Activity and Clinical Characteristics in Alcohol Use Disorders: Practical Issues and Results. *Clinical EEG and Neuroscience*. https://doi.org/10.1177/1550059415605686

Lempert, K. M., Speer, M. E., Delgado, M. R., & Phelps, E. A. (2017). Positive autobiographical memory retrieval reduces temporal discounting. *Social Cognitive and Affective Neuroscience*. https://doi.org/10.1093/scan/nsx086

Li, Z., Tong, L., Wang, L., Li, Y., He, W., Guan, M., & Yan, B. (2016). Self-regulating positive emotion networks by feedback of multiple emotional brain states using real-time fMRI. *Experimental Brain Research*. https://doi.org/10.1007/s00221-016-4744-z

Linhartová, P., Látalová, A., Kóša, B., Kašpárek, T., Schmahl, C., & Paret, C. (2019). fMRI neurofeedback in emotion regulation: A literature review. *NeuroImage*. https://doi.org/10.1016/j.neuroimage.2019.03.011

Lubar, J. F., Swartwood, M. O., Swartwood, J. N., & O'Donnell, P. H. (1995). Evaluation of the effectiveness of EEG neurofeedback training for ADHD in a clinical setting as measured by changes in T.O.V.A. scores, behavioral ratings, and WISC-R performance. *Biofeedback and Self-Regulation*. https://doi.org/10.1007/BF01712768

Maddock, R. J., Garrett, A. S., & Buonocore, M. H. (2001). Remembering familiar people: The posterior cingulate cortex and autobiographical memory retrieval. *Neuroscience*. https://doi.org/10.1016/S0306-4522(01)00108-7

Maldjian, J. A., Laurienti, P. J., & Burdette, J. H. (2004). Precentral gyrus discrepancy in electronic versions of the Talairach atlas. *NeuroImage*. https://doi.org/10.1016/j.neuroimage.2003.09.032

Maldjian, J. A., Laurienti, P. J., Kraft, R. A., & Burdette, J. H. (2003). An automated method for neuroanatomic and cytoarchitectonic atlas-based interrogation of fMRI data sets. *NeuroImage*. https://doi.org/10.1016/S1053-8119(03)00169-1

Marci, C. D., Glick, D. M., Loh, R., & Dougherty, D. D. (2007). Autonomic and prefrontal cortex responses to autobiographical recall of emotions. *Cognitive, Affective and*

*Behavioral Neuroscience*. https://doi.org/10.3758/CABN.7.3.243

Mayeli, A., Zotev, V., Refai, H., & Bodurka, J. (2016). Real-time EEG artifact correction during fMRI using ICA. *Journal of Neuroscience Methods*. https://doi.org/10.1016/j.jneumeth.2016.09.012

Megumi, F., Yamashita, A., Kawato, M., & Imamizu, H. (2015). Functional MRI neurofeedback training on connectivity between two regions induces long-lasting changes in intrinsic functional network. *Frontiers in Human Neuroscience*. https://doi.org/10.3389/fnhum.2015.00160

Mennella, R., Patron, E., & Palomba, D. (2017). Frontal alpha asymmetry neurofeedback for the reduction of negative affect and anxiety. *Behaviour Research and Therapy*. https://doi.org/10.1016/j.brat.2017.02.002

Meyer, T., Quaedflieg, C. W. E. M., Weijland, K., Schruers, K., Merckelbach, H., & Smeets, T. (2018). Frontal EEG asymmetry during symptom provocation predicts subjective responses to intrusions in survivors with and without PTSD. *Psychophysiology*. https://doi.org/10.1111/psyp.12779

Mitterschiffthaler, M. T., Fu, C. H. Y., Dalton, J. A., Andrew, C. M., & Williams, S. C. R. (2007). A functional MRI study of happy and sad affective states induced by classical music. *Human Brain Mapping*. https://doi.org/10.1002/hbm.20337

Mognon, A., Jovicich, J., Bruzzone, L., & Buiatti, M. (2011). ADJUST: An automatic EEG artifact detector based on the joint use of spatial and temporal features. *Psychophysiology*. https://doi.org/10.1111/j.1469-8986.2010.01061.x

Moosmann, M., Schönfelder, V. H., Specht, K., Scheeringa, R., Nordby, H., & Hugdahl, K. (2009). Realignment parameter-informed artefact correction for simultaneous EEG-fMRI recordings. *NeuroImage*. https://doi.org/10.1016/j.neuroimage.2009.01.024

Morawetz, C., Bode, S., Baudewig, J., & Heekeren, H. R. (2017). Effective amygdala-prefrontal connectivity predicts individual differences in successful emotion regulation. *Social Cognitive and Affective Neuroscience*. https://doi.org/10.1093/scan/nsw169

Niazy, R. K., Beckmann, C. F., Iannetti, G. D., Brady, J. M., & Smith, S. M. (2005). Removal of FMRI environment artifacts from EEG data using optimal basis sets. *NeuroImage*. https://doi.org/10.1016/j.neuroimage.2005.06.067

Nicholson, A. A., Rabellino, D., Densmore, M., Frewen, P. A., Paret, C., Kluetsch, R., … Lanius, R. A. (2017). The neurobiology of emotion regulation in posttraumatic stress disorder: Amygdala downregulation via real-time fMRI neurofeedback. *Human Brain Mapping*. https://doi.org/10.1002/hbm.23402

Ochsner, K. N., & Gross, J. J. (2008). Cognitive Emotion Regulation: Insights From Social Cognitive and Affective Neuroscience. *Current Directions in Psychological Science*. https://doi.org/10.1111/j.1467-8721.2008.00566.x

Paret, C., Kluetsch, R., Zaehringer, J., Ruf, M., Demirakca, T., Bohus, M., … Schmahl, C. (2016). Alterations of amygdala-prefrontal connectivity with real-time fMRI neurofeedback in BPD patients. *Social Cognitive and Affective Neuroscience*. https://doi.org/10.1093/scan/nsw016

Paret, C., Ruf, M., Gerchen, M. F., Kluetsch, R., Demirakca, T., Jungkunz, M., … Ende, G. (2016). FMRI neurofeedback of amygdala response to aversive stimuli enhances prefrontal-limbic brain connectivity. *NeuroImage*. https://doi.org/10.1016/j.neuroimage.2015.10.027

Peeters, F., Oehlen, M., Ronner, J., Van Os, J., & Lousberg, R. (2014). Neurofeedback As a Treatment for Major Depressive Disorder -A Pilot Study. *PLoS ONE*. https://doi.org/10.1371/journal.pone.0091837

Pelletier, M., Bouthillier, A., Lévesque, J., Carrier, S., Breault, C., Paquette, V., … Beauregard, M. (2003). Separate neural circuits for primary emotions? Brain activity during self-induced sadness and happiness in professional actors. *NeuroReport*. https://doi.org/10.1097/00001756-200306110-00003

Phelps, E. A. (2004). Human emotion and memory: Interactions of the amygdala and hippocampal complex. *Current Opinion in Neurobiology*. https://doi.org/10.1016/j.conb.2004.03.015

Pohl, A., Anders, S., Schulte-Rüther, M., Mathiak, K., & Kircher, T. (2013). Positive Facial Affect - An fMRI Study on the Involvement of Insula and Amygdala. *PLoS ONE*. https://doi.org/10.1371/journal.pone.0069886

Quaedflieg, C. W. E. M., Smulders, F. T. Y., Meyer, T., Peeters, F., Merckelbach, H., & Smeets, T. (2015). The validity of individual frontal alpha asymmetry EEG neurofeedback. *Social Cognitive and Affective Neuroscience*. https://doi.org/10.1093/scan/nsv090

Ros, T., Théberge, J., Frewen, P. A., Kluetsch, R., Densmore, M., Calhoun, V. D., & Lanius, R. A. (2013). Mind over chatter: Plastic up-regulation of the fMRI salience network directly after EEG neurofeedback. *NeuroImage*. https://doi.org/10.1016/j.neuroimage.2012.09.046

Rossi-Izquierdo, M., Ernst, A., Soto-Varela, A., Santos-Perez, S., Faraldo-Garcia, A., Sesar-Ignacio, A., & Basta, D. (2013). Vibrotactile neurofeedback balance training in patients with Parkinson's disease: reducing the number of falls. *Gait & Posture*. https://doi.org/10.1016/j.gaitpost.2012.07.002

Sarkheil, P., Zilverstand, A., Kilian-Hütten, N., Schneider, F., Goebel, R., & Mathiak, K. (2015). fMRI feedback enhances emotion regulation as evidenced by a reduced amygdala response. *Behavioural Brain Research*. https://doi.org/10.1016/j.bbr.2014.11.027

Saxby, E., & Peniston, E. G. (1995). Alpha-theta brainwave neurofeedback training: An effective treatment for male and female alcoholics with depressive symptoms. *Journal of Clinical Psychology*. https://doi.org/10.1002/1097-4679(199509)51:5<685::AID-JCLP2270510514>3.0.CO;2-K

Schutter, D. J., & van Honk, J. (2006). Increased positive emotional memory after repetitive transcranial magnetic stimulation over the orbitofrontal cortex. *Journal of Psychiatry and Neuroscience*.

Sherwood, M. S., Kane, J. H., Weisend, M. P., & Parker, J. G. (2016). Enhanced control of dorsolateral prefrontal cortex neurophysiology with real-time functional magnetic resonance imaging (rt-fMRI) neurofeedback training and working memory practice. *NeuroImage*. https://doi.org/10.1016/j.neuroimage.2015.08.074

Smith, S. M., Jenkinson, M., Woolrich, M. W., Beckmann, C. F., Behrens, T. E. J., Johansen-Berg, H., … Matthews, P. M. (2004). Advances in functional and structural MR image analysis and implementation as FSL. In *NeuroImage*. https://doi.org/10.1016/j.neuroimage.2004.07.051

Spielberger, C. D. (1972). Profile of Mood States. *Professional Psychology*. https://doi.org/10.1037/h0020742

Suardi, A., Sotgiu, I., Costa, T., Cauda, F., & Rusconi, M. (2016). The neural correlates of happiness: A review of PET and fMRI studies using autobiographical recall methods. *Cognitive, Affective and Behavioral Neuroscience*. https://doi.org/10.3758/s13415-016-0414-7

Subramanian, L., Hindle, J. V., Johnston, S., Roberts, M. V., Husain, M., Goebel, R., & Linden, D. (2011). Real-Time Functional Magnetic Resonance Imaging Neurofeedback for Treatment of Parkinson's Disease. *The Journal of Neuroscience*. https://doi.org/10.1523/JNEUROSCI.3498-11.2011

Sulzer, J., Haller, S., Scharnowski, F., Weiskopf, N., Birbaumer, N., Blefari, M. L., … Sitaram, R. (2013). Real-time fMRI neurofeedback: Progress and challenges. *NeuroImage*. https://doi.org/10.1016/j.neuroimage.2013.03.033

Svoboda, E., McKinnon, M. C., & Levine, B. (2006). The functional neuroanatomy of autobiographical memory: A meta-analysis. *Neuropsychologia*. https://doi.org/10.1016/j.neuropsychologia.2006.05.023

Tzourio-Mazoyer, N., Landeau, B., Papathanassiou, D., Crivello, F., Etard, O., Delcroix, N., … Joliot, M. (2002). Automated anatomical labeling of activations in SPM using a macroscopic anatomical parcellation of the MNI MRI single-subject brain. *NeuroImage*. https://doi.org/10.1006/nimg.2001.0978

Veit, R., Singh, V., Sitaram, R., Caria, A., Rauss, K., & Birbaumer, N. (2012). Using real-time fmri to learn voluntary regulation of the anterior insula in the presence of threat-related stimuli. *Social Cognitive and Affective Neuroscience*. https://doi.org/10.1093/scan/nsr061

Vrticka, P., Simioni, S., Fornari, E., Schluep, M., Vuilleumier, P., & Sander, D. (2013). Neural substrates of social emotion regulation: A fMRI study on imitation and expressive suppression to dynamic facial signals. *Frontiers in Psychology*. https://doi.org/10.3389/fpsyg.2013.00095

Walker, J. E., & Kozlowski, G. P. (2005). Neurofeedback treatment of epilepsy. *Child and Adolescent Psychiatric Clinics of North America*. https://doi.org/10.1016/j.chc.2004.07.009

Watson, D., Clark, L. A., & Tellegen, A. (1988). Development and Validation of Brief Measures of Positive and Negative Affect: The PANAS Scales. *Journal of Personality and Social Psychology*. https://doi.org/10.1037/0022-3514.54.6.1063

Wong, C. K., Zotev, V., Misaki, M., Phillips, R., Luo, Q., & Bodurka, J. (2016). Automatic EEG-assisted retrospective motion correction for fMRI (aE-REMCOR). *NeuroImage*. https://doi.org/10.1016/j.neuroimage.2016.01.042

Woolrich, M. W., Jbabdi, S., Patenaude, B., Chappell, M., Makni, S., Behrens, T., … Smith, S. M. (2009). Bayesian analysis of neuroimaging data in FSL. *NeuroImage*. https://doi.org/10.1016/j.neuroimage.2008.10.055

Young, K. D., Siegle, G. J., Misaki, M., Zotev, V., Phillips, R., Drevets, W. C., & Bodurka, J. (2018). Altered task-based and resting-state amygdala functional connectivity following real-time fMRI amygdala neurofeedback training in major depressive disorder. *NeuroImage: Clinical*. https://doi.org/10.1016/j.nicl.2017.12.004

Young, K. D., Zotev, V., Phillips, R., Misaki, M., Yuan, H., Drevets, W. C., & Bodurka, J. (2014). Real-time fMRI neurofeedback training of amygdala activity in patients with major depressive disorder. *PLoS ONE*. https://doi.org/10.1371/journal.pone.0088785

Zotev, V, Yuan, H., Phillips, R., & Bodurka, J. (2012). EEG-assisted retrospective motion correction for fMRI: E-REMCOR. *Arxiv Preprint ArXiv:1201.4481*. https://doi.org/10.1016/j.neuroimage.2012.07.031

Zotev, Vadim, Krueger, F., Phillips, R., Alvarez, R. P., Simmons, W. K., Bellgowan, P., … Bodurka, J. (2011a). Self-regulation of amygdala activation using real-time FMRI neurofeedback. *PLoS ONE*. https://doi.org/10.1371/journal.pone.0024522

Zotev, Vadim, Krueger, F., Phillips, R., Alvarez, R. P., Simmons, W. K., Bellgowan, P., … Bodurka, J. (2011b). Self-regulation of amygdala activation using real-time FMRI neurofeedback. *PLoS ONE*. https://doi.org/10.1371/journal.pone.0024522

Zotev, Vadim, Misaki, M., Phillips, R., Wong, C. K., & Bodurka, J. (2018). Real-time fMRI neurofeedback of the mediodorsal and anterior thalamus enhances correlation between thalamic BOLD activity and alpha EEG rhythm. *Human Brain Mapping*. https://doi.org/10.1002/hbm.23902

Zotev, Vadim, Phillips, R., Yuan, H., Misaki, M., & Bodurka, J. (2014). Self-regulation of human brain activity using simultaneous real-time fMRI and EEG neurofeedback. *NeuroImage*. https://doi.org/10.1016/j.neuroimage.2013.04.126

Zotev, Vadim, Yuan, H., Misaki, M., Phillips, R., Young, K. D., Feldner, M. T., & Bodurka, J. (2016). Correlation between amygdala BOLD activity and frontal EEG asymmetry during real-time fMRI neurofeedback training in patients with depression. *NeuroImage: Clinical*. https://doi.org/10.1016/j.nicl.2016.02.003

Zuberer, A., Minder, F., Brandeis, D., & Drechsler, R. (2018). Mixed-Effects Modeling of Neurofeedback Self-Regulation Performance: Moderators for Learning in Children with ADHD. *Neural Plasticity*. https://doi.org/10.1155/2018/2464310

## Tables

Table 1: Percentages of signal change, t-score, and Cohen's D effect size for brain regions with significant signal change between Upregulation versus View and Rest in the experimental group (FDR-corrected for multiple comparison, q = 0.01).

| Regions | Sig % UP-View | Sig % Up-Rest | t-score (UP-View) - MNI coordinate | Cohen's D effect size (UP-View) | Sig % Other researches (Up-Rest) |
|---|---|---|---|---|---|
| Left Amygdala | 0.86 | 0.70 | 4.9 (-24,0,-12) | 0.87 | 0.7 (Young et al., 2014), 0.3, 0.1 (Li et al., 2016), 0.2 (D. Y. Kim et al., 2015) |
| Right Amygdala | 0.65 | 0.72 | 3.9 (24,6,-16) | 0.70 | 0.4 (Young et al., 2014), 0.3 (Vadim Zotev et al., 2016) |
| Left Insula | 1 | 0.64 | 7.7 (-44,4,-4) | 0.85 | 0.5 (S. J. Johnston et al., 2010), 0.5 (Li et al., 2016) |
| Right Insula | 0.91 | 0.62 | 6.4 (46,6,4) | 0.88 | - |
| Left Anterior Cingulate Cortex | 0.97 | 0.81 | 4.2 (-6,12,28) | 0.95 | 0.3 (Li et al., 2016) |
| Right Anterior Cingulate Cortex | 0.64 | 0.38 | 4.4 (18,44,14) | 0.86 | |
| Left Cuneus | 0.45 | 1.56 | 4.5 (-20,-56,26) | 0.61 | 0.5 (S. J. Johnston et al., 2010) |
| Right Cuneus | 0.40 | 1.90 | 3.9 (26,-62,26) | 0.75 | - |
| Left Lingual Gyrus | 1.21 | 1.39 | 4.2 (-12,-68,-10) | 0.83 | - |
| Left Posterior Cingulate Cortex | 0.49 | 0.33 | 5 (-14,-52,58) | 0.66 | 0.5 (S. J. Johnston et al., 2010) |
| Left Thalamus | 1.07 | 0.85 | 4.9 (-22,-24,6) | 0.93 | - |
| Right Thalamus | 0.86 | 0.65 | 6.9 (18,-16,18) | 0.90 | - |
| Left Caudate | 0.86 | 0.65 | 5.7 (-16,12,10) | 0.80 | - |
| Right Caudate | 0.74 | 0.49 | 8 (18,-14,20) | 0.80 | - |
| Left Hippocampus | 0.57 | 0.56 | 4.3 (-36,-18,-12) | 0.85 | |
| Right Hippocampus | 0.44 | 0.59 | 4.2 (40,-26,-10) | 0.80 | - |
| Left Dorsomedial Prefrontal Cortex | 0.85 | 1.02 | 5.1 (0,20,42) | 0.99 | |
| Right Dorsomedial Prefrontal Cortex | 0.37 | 0.81 | 4.1 (10,24,42) | 0.89 | - |
| Left Orbitofrontal Cortex | 1.13 | 1.04 | 6 (-44,18,-2) | 1 | - |
| Right Orbitofrontal Cortex | 1.12 | 0.81 | 6.5 (30,32,-12) | 0.85 | - |
| Left Middle Temporal Gyrus | 0.66 | 0.59 | 5.9 (-48,-30,-26) | 0.87 | |
| Right Middle Temporal Gyrus | 0.69 | 0.70 | 5.9 (56,-20,-28) | 0.70 | - |
| Left Ventral Striatum | 1.17 | 0.84 | 6.1 (-16,19,-1) | 1.03 | 0.5 (S. J. Johnston et al., 2010) |
| Right Ventral Striatum | 0.81 | 0.66 | 7.7 (19,16,-2) | 0.65 | 0.5 (S. J. Johnston et al., 2010) |
| Left Ventrolateral Prefrontal Cortex | 0.67 | 0.81 | 5.6 (-46,18,2) | 0.82 | 0.5 (S. J. Johnston et al., 2010) |
| Right Ventrolateral Prefrontal Cortex | 0.65 | 0.58 | 9.3 (62,20,0) | 0.60 | - |
| Left Dorsolateral Prefrontal Cortex | 0.84 | 0.90 | 5.3 (-28,44,18) | 0.97 | - |
| Right Dorsolateral Prefrontal Cortex | 0.76 | 0.75 | 6.7 (40,-2,52) | 0.88 | - |
| Left Superior Parietal | 0.46 | 0.69 | 4.9 (-24,-52,50) | 0.81 | - |
| Right Superior Parietal | 0.33 | 0.89 | 5.3 (16,-48,54) | 0.57 | - |
| Left Inferior Parietal | 0.60 | 0.41 | 5.9 (-60,-36,44) | 0.85 | - |
| Right Inferior Parietal | 1.32 | 0.53 | 4.2 (56,-36,54) | 1.01 | - |
| Left SupraMarginal | 0.87 | 0.42 | 8.1 (-54,-38,24) | 1.04 | - |
| Left Postcentral | 0.67 | 0.56 | 6.8 (-50,-24,32) | 0.88 | - |

Table 2: Psychometric test results before and after neurofeedback for the experimental group (changes from before to after neurofeedback for positive and negative mood states, POMS and TMD are significant (paired t-test; df=17).

| Measure | scores before neurofeedback | scores after neurofeedback | Effect size (*d*) | *p*-value |
|---|---|---|---|---|
| PANAS | 52.2 ± 11.5 | 51.6 ± 8.8 | -0.058 | 0.7 |
| PANAS negative mood states | 20.8 ± 7.2 | 14.1 ± 4.8 | -1.09 | $2 \times 10^{-3}$ |
| PANAS positive mood states | 31.4 ± 6.1 | 37.5 ± 6.4 | 0.97 | $4.5 \times 10^{-4}$ |
| POMS | 24.6 ± 10.9 | 17 ± 6.9 | -0.83 | $1.1 \times 10^{-3}$ |
| Total Mood Distribution (TMD) | 7.5 ± 11.5 | -4.7 ± 7.1 | -1.27 | $2.1 \times 10^{-4}$ |

Table 3: Psychometric test results before and after neurofeedback for the control group (changes from before to after neurofeedback are significant for none of the psychometric tests).

| Measure | scores before neurofeedback | scores after neurofeedback | Effect size (*d*) | *p*-value |
|---|---|---|---|---|
| PANAS | 54.2 ± 5.9 | 53 ± 4.8 | -0.22 | 0.1 |
| PANAS negative mood states | 22.1 ± 6 | 19.3 ± 5 | -0.50 | 0.064 |
| PANAS positive mood states | 32.1 ± 5.7 | 33.7 ± 5.6 | 0.28 | 0.33 |
| POMS | 27.1 ± 11.3 | 22.8 ± 11.1 | -0.38 | 0.08 |
| Total Mood Distribution (TMD) | 6.6 ± 11.2 | 3.6 ± 11.7 | -0.26 | 0.15 |

Table 4: Significant connectivity obtained in previous emotion regulation researches and this study as a result of neurofeedback.

| Study # | References | Brain area |
|---|---|---|
| 1 | (Young et al., 2018) | amygdala and:<br>inferior frontal G/lateral, medial PFC, medial frontopolar Cortex, ventrolateral PFC, medial frontal Gyrus, ACC, insula, ventral striatum, putamen, thalamus, precuneus, cerebellum, temporal pole |
| 2 | (Paret, Ruf, et al., 2016) | amygdala and: supplementary motor area, middle frontal gurus, brain stem, precuneus, ventromedial prefrontal cortex, white matter/ right putamen/insula |
| 3 | (Vadim Zotev et al., 2011b) | amygdala and : frontal lobe (Ventrolateral prefrontal cortex, orbitofrontal cortex, Middle frontal gyrus, superior frontal gyrus, ventromedial prefrontal cortex), temporal Lobe (middle temporal gyrus), limbic Lobe (hippocampus), sub-lobar Regions (insula, thalamus) |
| 4 | (Paret, Kluetsch, et al., 2016) | amygdala and: DLPFC, Precentral gyrus, paracentral lobe, parahippocampal gyrus, extending to thalamus and hippocampus, cerebellum |
| 5 | (Herwig et al., 2019) | amygdala and: ACC, DLPFC, DMPFC, pre-SMA and VLPFC |
| 6 | (Koush et al., 2017) | amygdala-DLPFC |
| 7 | (Nicholson et al., 2017) | amygdala and: DMPFC, DLPFC, ACC |
| 8 | (Sarkheil et al., 2015) | PFC-PCC |
| 9 | (Veit et al., 2012) | insula and: lingual gyrus, ventrolateral PFC, Frontal inferior operculum, inferior orbitofrontal, Middle frontal, Middle orbitofrontal, Occipital inferior, Dorsal medial PFC, |
| 10 | (Cohen Kadosh et al., 2016) | insula and: mid cingulate cortex, supplementary motor area, amygdala |
| 11 | (Ros et al., 2013) | ACC and mid-cingulate cortex. |
| 12 | (Banks, Eddy, Angstadt, Nathan, & | amygdala and: OFC and DMPFC |
| | Luan Phan, 2007) | |
| 13 | (Morawetz, Bode, Baudewig, & Heekeren, 2017) | inferior frontal gyrus (IFG) and: DLPFC, DMPFC, VMPFC, ACC, amygdala<br>amygdala and prefrontal regions (e.g. DMPFC) |
| 14 | (HE et al., 2017) | amygdala and: ACC amd thalamus<br>PCC and centromedial amygdala (subregions of amygdala) |
| 15 | (Koush et al., 2019) | temporoparietal junction (TPJ) and: SFG, DMPFC, vmPFC<br>vmPFC and: DMPFC, amygdala, |
| 16 | This study | **Connections obtained in this study and reported in previous studies:**<br>amygdala – thalamus, amygdala - dorsomedial prefrontal cortex, thalamus – ventral striatum, insula - dorsomedial prefrontal cortex<br>**New connections:**<br>thalamus - dorsomedial prefrontal cortex, thalamus – insula, thalamus - orbitofrontal cortex, thalamus - ventrolateral prefrontal cortex, thalamus – postcentral, dorsomedial prefrontal cortex – ventral striatum, orbitofrontal cortex - ventrolateral prefrontal cortex |

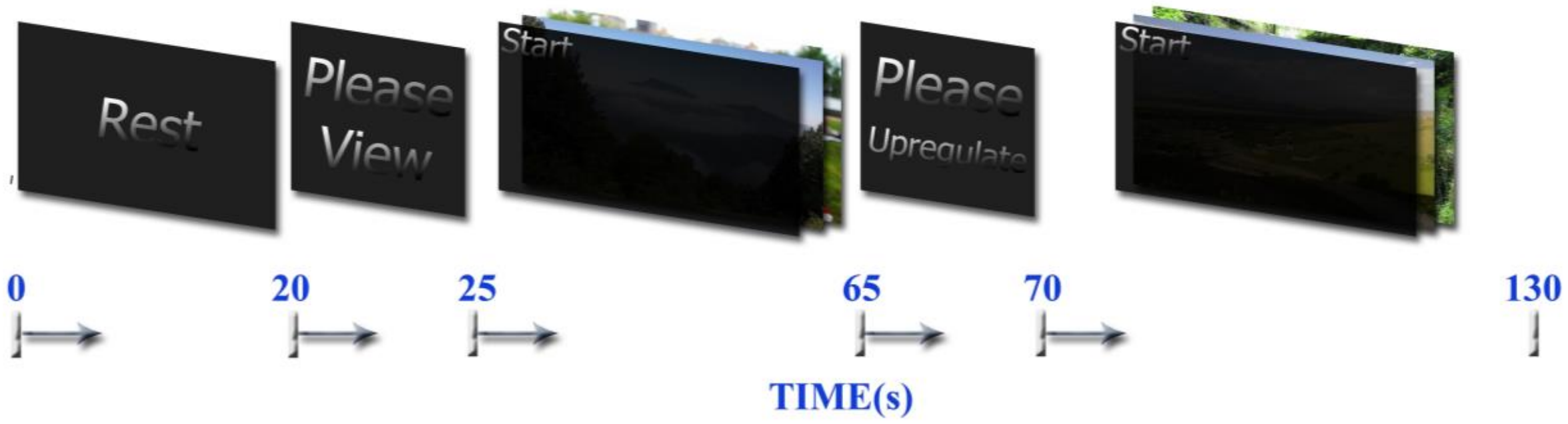

Figure 1: Timing of the tasks used in the one run of neurofeedback protocol, which contains rest, view, and upregulation blocks.

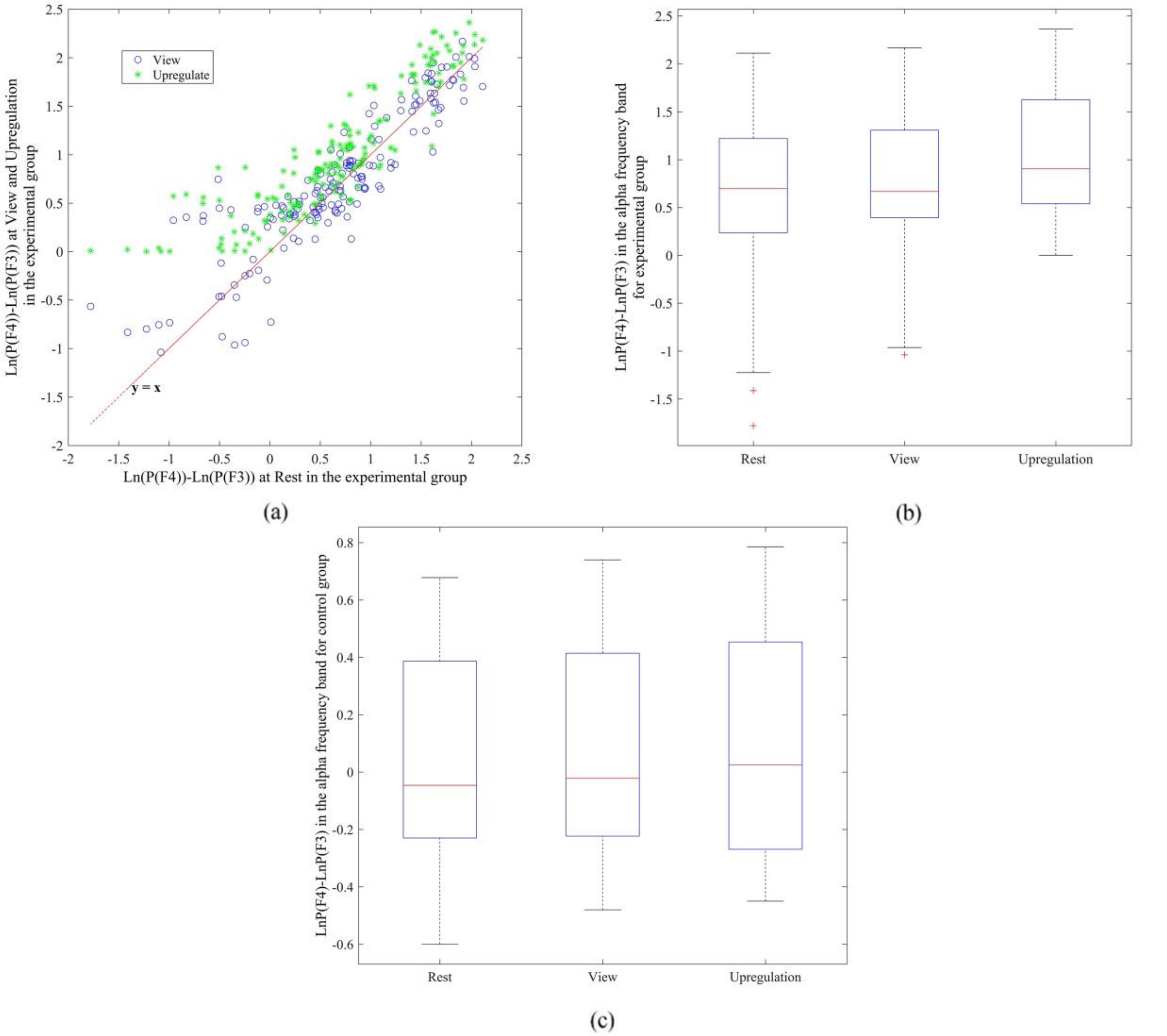

Figure 2: (a) Scatter plot of frontal asymmetry in blocks of View and Upregulation versus Rest for the experimental group. (b) Boxplot of frontal asymmetry in blocks of Rest, View, and Upregulation for the experimental group. (c) Boxplot of frontal asymmetry in blocks of Rest, View, and Upregulation for the control group.

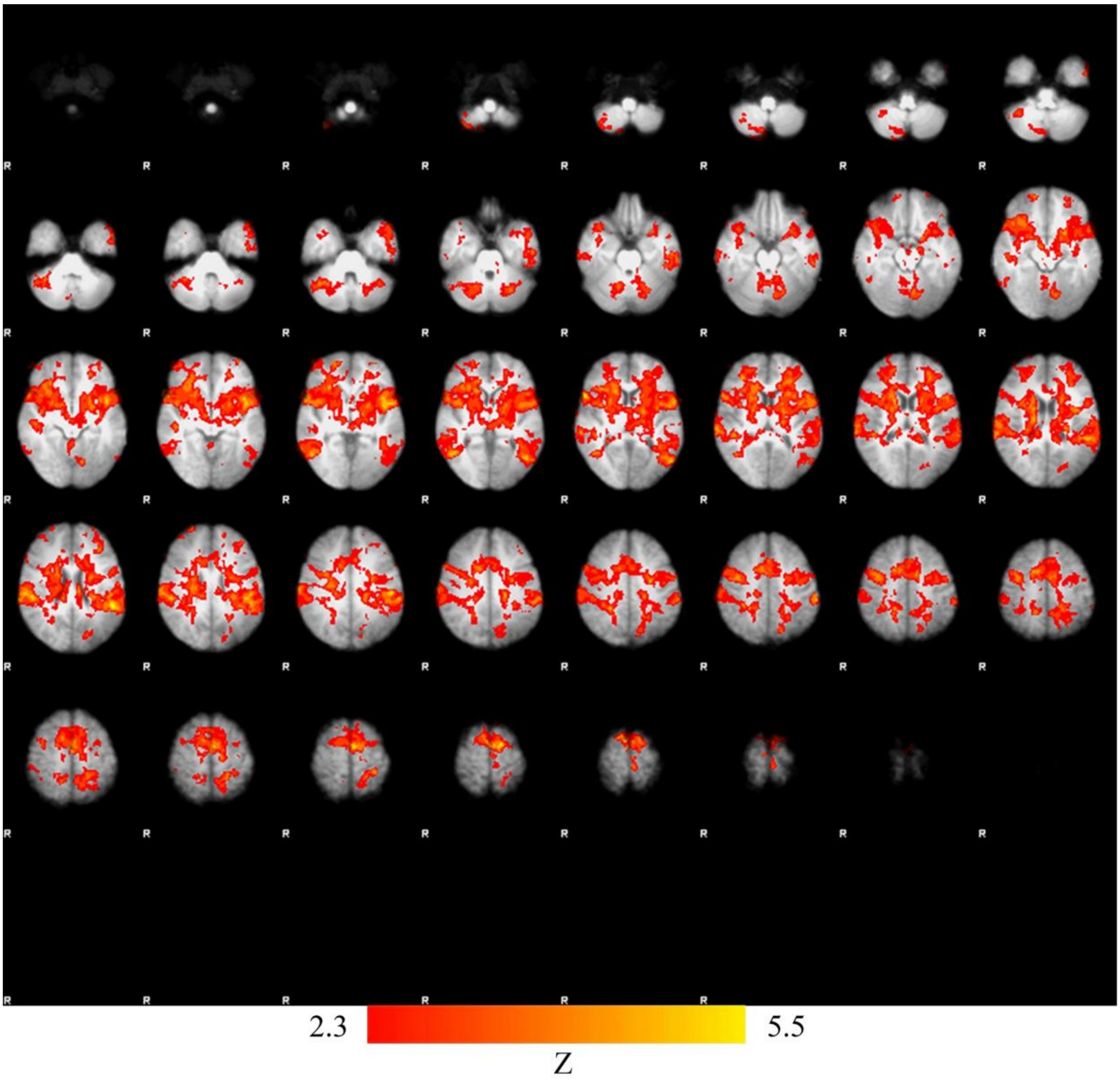

Figure 3: Activation map for Upregulation versus View in the experimental group ($p_{corrected}$ = 0.01).

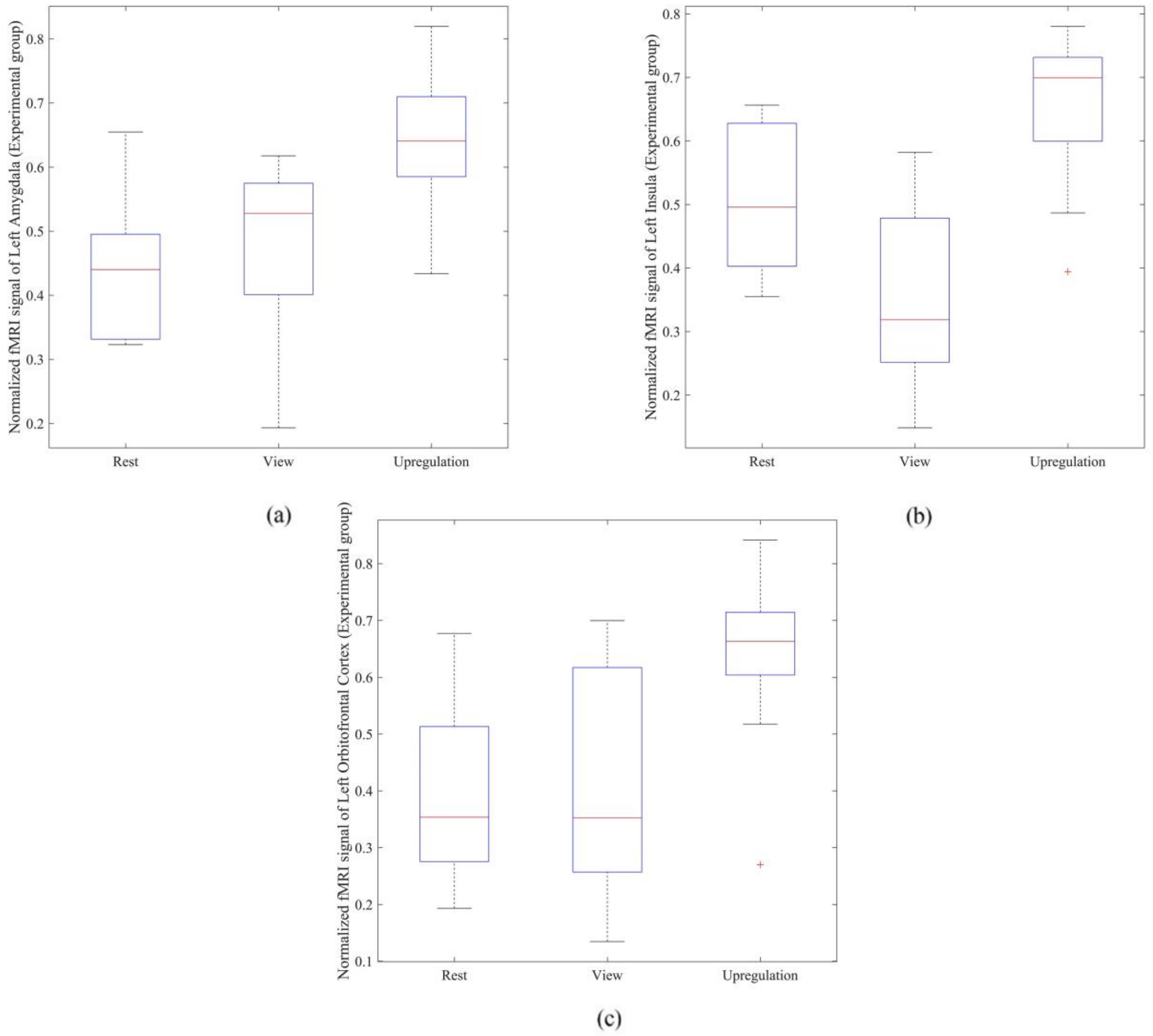

Figure 4: Normalized fMRI BOLD signal in the Rest, View, and Upregulation blocks for contrast of Upregulation versus View and Rest for the experimental group in: a) left Amygdala; b) left Orbitofrontal Cortex; and c) left Insula.

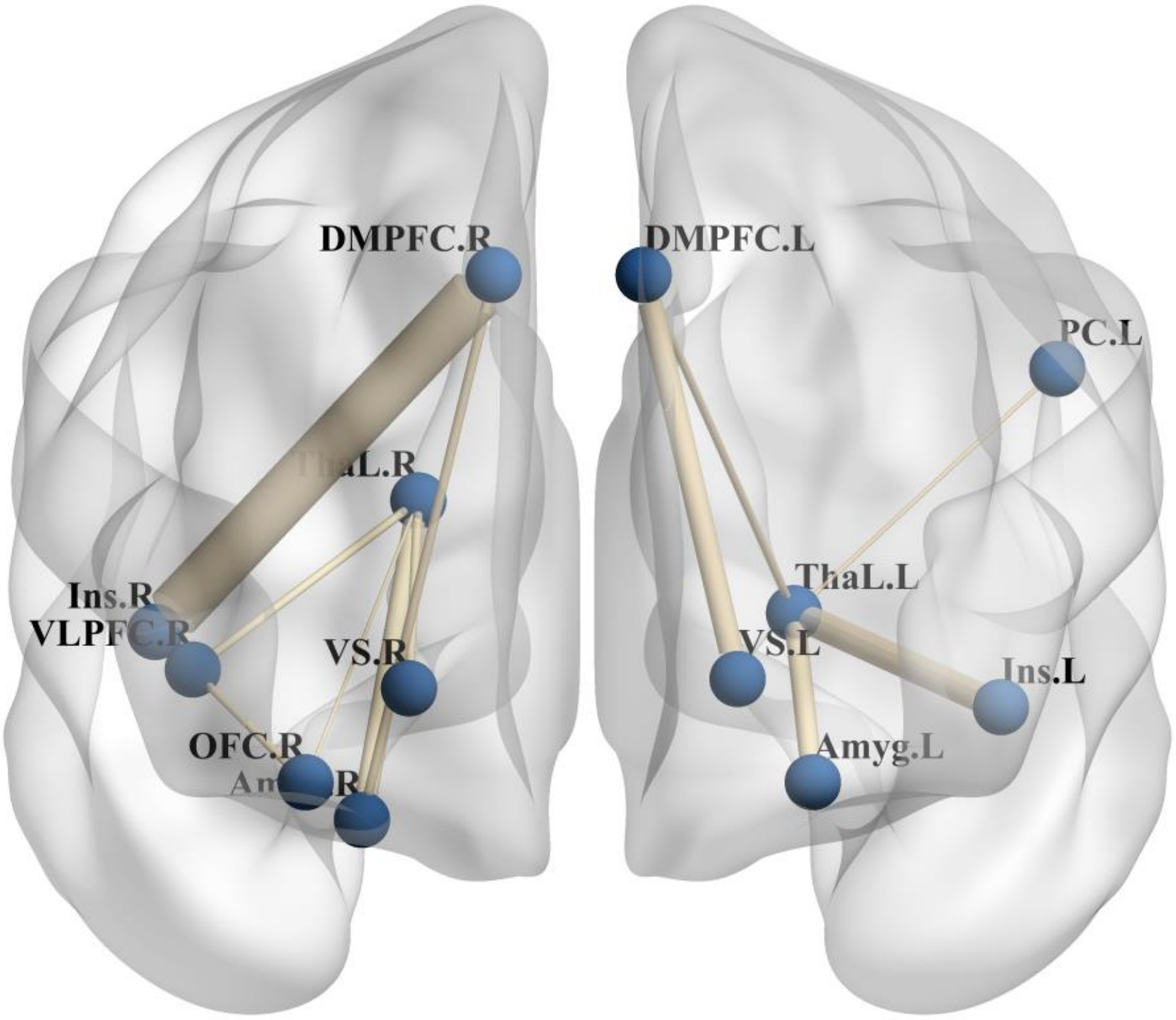

Figure 5: Significant edges of differential connectivity networks which show network links with significant changes between the Upregulation and View blocks. L, left; R, right; Amyg, Amygdala; ThaL, thalamus; Ins, insula; OFC, orbitofrontal cortex; VS, ventral striatum; DMPFC, dorsomedial prefrontal cortex; VLPFC, ventrolateral prefrontal cortex; PC, postcentral.

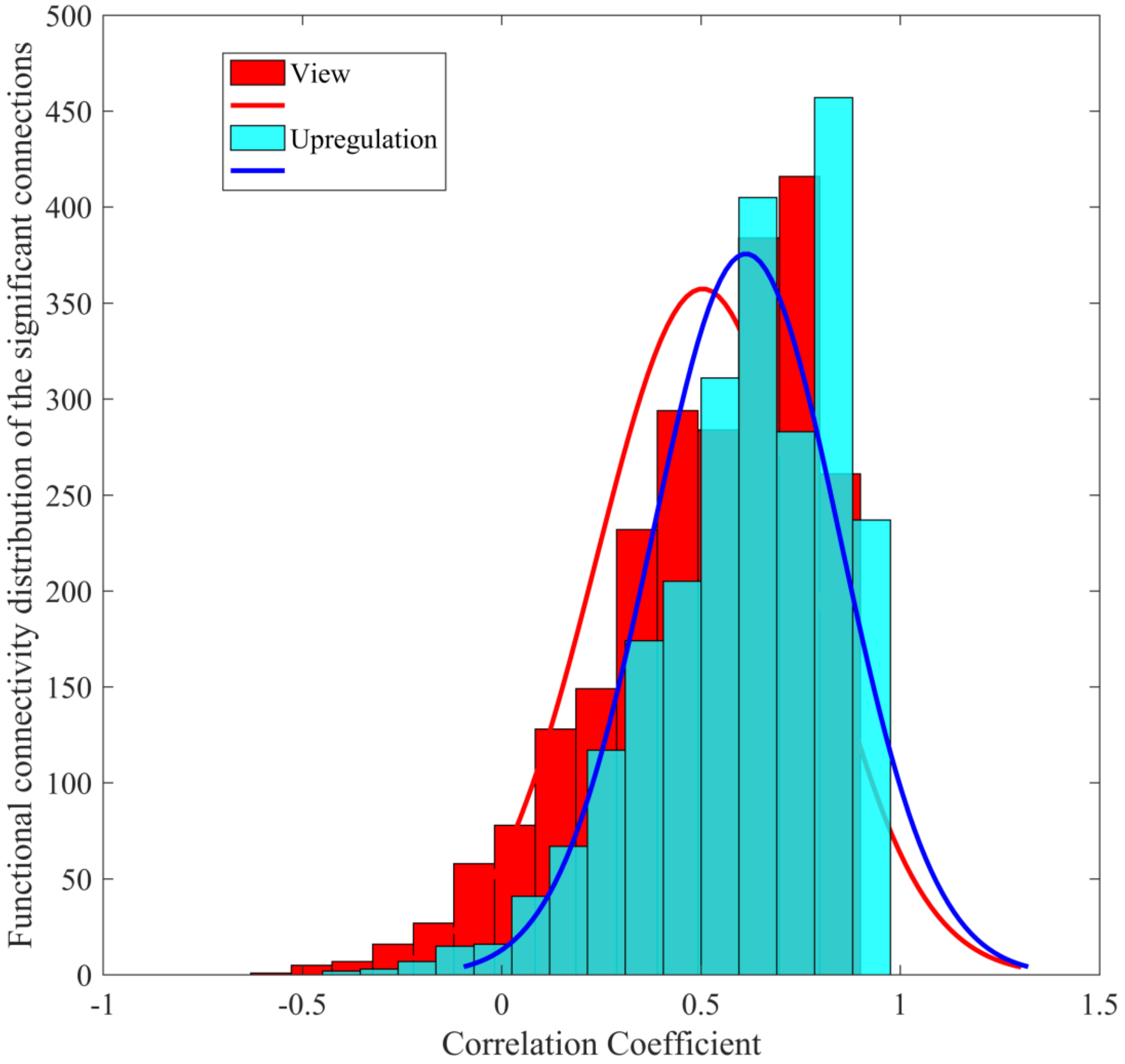

Figure 6: Functional connectivity distribution of upregulation and view blocks for significant connections in the experimental group.